%% file: firstyearproject.tex
\documentclass[a4paper,nobind]{ociamthesis}

\correctionstrue

\title{Out-of-time correlation functions in single-body systems}
\author{\textbf{Andrew C. Hunt}}
\college{Caius College}
\degreedate{First-Year Report}

\usepackage{import}
\usepackage{amsmath}
\usepackage{physics}
\usepackage{graphicx}
\usepackage{mathrsfs}
\usepackage{siunitx}
\usepackage[section]{placeins}
\usepackage{upgreek}
\usepackage{dsfont}
\usepackage{caption}
\usepackage{subcaption}
\usepackage{multirow}
\usepackage{hhline}
\usepackage{relsize}
\usepackage{mathdots}
\usepackage{cite}
\usepackage{booktabs}
\usepackage{float}
\usepackage{tikz}
\usepackage{xcolor}

	\renewcommand{\th}{\textsuperscript{th}} 

    \newcommand{\vc}{\boldsymbol}
    \newcommand{\mt}{\boldsymbol}
    \newcommand{\1}{\hat{\mathds{1}}}
    
    \newcommand{\etal}{\textit{et al.\:}}
    \newcommand{\novtwo}{\frac{N}{2}}
    \renewcommand{\d}{{\rm d}} 

\newcommand{\mathcolorbox}[2]{\mathchoice
  {\colorbox{#1}{$\displaystyle #2$}}
  {\colorbox{#1}{$\textstyle #2$}}
  {\colorbox{#1}{$\scriptstyle #2$}}
  {\colorbox{#1}{$\scriptscriptstyle #2$}}
}

\begin{document}

\setlength{\textbaselineskip}{17pt plus2pt minus1pt}

\setlength{\frontmatterbaselineskip}{17pt plus1pt minus1pt}

\setlength{\baselineskip}{\textbaselineskip}

\setlength{\abovedisplayskip}{14pt plus3pt minus9pt}
\setlength{\belowdisplayskip}{14pt plus3pt minus9pt}


\setcounter{secnumdepth}{2}
\setcounter{tocdepth}{2}

\begin{romanpages}

\maketitle



\begin{abstract}
	\input{text/abstract}
\end{abstract}

\dominitoc 

\flushbottom

\tableofcontents






\end{romanpages}

\include{text/C0-Intro/intro}
\include{text/C1/C1}
\include{text/C2/C2}
\include{text/C2/C2+1}

\include{text/C3/C3}

\include{text/C4/Conclusion}


{\footnotesize
\setlength{\baselineskip}{0pt}
\renewcommand*\MakeUppercase[1]{#1}
\bibliography{bibliography/refs}{}}


\startappendices
\include{text/CounterExParameters}

\include{text/A-Shake}
\include{text/MeomMatsDerivation}
\include{text/ITTIstationaryProof}

\include{text/WavepacketCalculation}



\end{document}

%% file: text/abstract.tex
In the study of quantum chaos, `out of time ordered correlators' (OTOCs) are commonly used to quantify the rate at which quantum information is scrambled. This rate
has been conjectured by Maldecena \etal to obey a universal, temperature dependent bound.
Recent studies have shown that instantons, delocalised structures that dominate tunnelling statistics over barriers, reduce the growth rate of OTOCs. For the case of the symmetric double well, this reduction ensures the bound is maintained for OTOCs generated using ring polymer molecular dynamics (RPMD), a method with approximate dynamics but exact quantum statistics. In this report we set out to further understand the role of the instanton in the enforcement of the Maldacena bound and test whether RPMD is sufficient to satisfy the bound. We also investigate the impact of coherence on the flattening of of OTOCs by contrasting bounded with scattering systems. For the scattering system we observe a significantly smaller OTOC growth rate than that of the analogous bounded system, and a flattening in growth rate as time progresses. We attribute the first effect to influence of the Boltzmann operator, and the second to interference caused by anharmonicity of the potential, but require further for verification of these conclusions. In our studies of RPMD, we find counterexamples showing that it is not sufficient to satisfy the bound. In light of these results, we develop a theory for OTOCs using (analytically-continued) Matsubara dynamics, revealing significantly different dynamical behaviour around the instanton compared to the predictions of RPMD. 
The instanton is found to be stationary in all coordinates but its collective angle $\Phi_0$, and fluctuations about it no longer resemble that of classical dynamics on a first order saddle as in RPMD. Advancing this theory could further help elucidate the origin of the Maldacena bound and serves as a starting point in the development of novel instanton-based quantum rate theories. 

%% file: text/C0-Intro/intro.tex
\chapter{\label{1:}Introduction} 
Classical chaos investigates the complex and unpredictable behaviours of dynamical systems. It is observed in a vast range of systems, from the dynamics of biological systems\cite{olsenChaosBiologicalSystems1985} to the orbits of celestial objects\cite{poincareMethodesNouvellesMecanique1899} and is the cornerstone on which the ergodic hypothesis of classical statistical mechanics lies\cite{buzziChaosErgodicTheory2009}. A commonly used diagnostic of classical chaos is `exponential sensitivity to small changes in initial conditions', resulting in the `butterfly effect'.
\\[5pt]
\indent 
The Heisenberg uncertainty principle forbids the specification of both position and momentum simultaneously in quantum systems. This presents a challenge for an analogous `butterfly effect' like description of quantum chaos, as initial conditions can not all be specified. In addition to this, due to the linearity of the time-dependent Schr\"odinger equation, chaotic behaviour analogous to that of the corresponding classical phase space variables cannot arise in the expectation values of Hermitian operators.
For these reasons, the concept of `quantum chaos' analogous to the classical case does not exist. Instead, quantum chaos refers to the study of quantized systems whose classical counterparts exhibit chaotic behaviour and the investigation of how the localisation of initial wavefunctions spreads throughout quantum systems over time, also known as `scrambling'.
\\[5pt]
\indent 
`Out-of-time ordered correlation functions' (OTOCs) were originally introduced by Larkin and Ovchinnikov in their discussion of semiclassical approaches to superconductivity\cite{larkinQuasiclassicalMethodTheory1969}. Following this, OTOCs have been reintroduced and extensively studied\footnotemark{}\footnotetext{There are far too many citations to reference here. For a comprehensive review see Ref.\cite{garcia-mataOutoftimeorderCorrelationsQuantum2023}.} as a measure of quantum scrambling. Their exponential growth rate, commonly referred to as the `quantum  Lyapunov exponent', is conjectured to obey a sharp, system independent bound proposed by Maldacena \etal\!\!\!\cite{maldacenaBoundChaos2016}, setting a temperature-dependent `speed limit' for quantum scrambling. Given the universal nature of the bound, uncovering its origins could provide insights into new fundamental properties of quantum dynamics.
\\[5pt]
\indent
The Maldacena bound has been hypothesised by multiple authors \cite{murthyBoundsChaosEigenstate2019a,tsujiBoundExponentialGrowth2018,pappalardiQuantumBoundsFluctuationdissipation2022a,nussinovExactUniversalChaos2022} to be statistical in origin. In a recent study, Sadhasivam \etal demonstrated,\cite{sadhasivamInstantonsQuantumBound2023} that ring polymer molecular dynamics (RPMD), an approximate method based on the path integral formulation of quantum mechanics that conserves the quantum Boltzmann distribution\cite{craigRingPolymerMolecular2006}, generated OTOCs that strictly adhered to the Maldacena bound in the example of a symmetric double well. 
\\[5pt]
\indent 
We aim to address two main questions in this report. The first is to determine whether RPMD is a sufficient method to satisfy the bound and understand the mechanism by which it is maintained. The second is to understand the role of coherence in the growth rate of OTOCs. 
\\[5pt]
\indent 
The report begins with a review of result of Sadhasivam \etal and the theory behind OTOCs. In Chapters~\ref{3:} and~\ref{4:} we test this hypothesis and expand upon it using a Matsubara dynamics-based approach. Chapter~\ref{5:} focuses on the behaviour of quantum OTOCs in scattering systems. We then conclude with a summary of our findings and potential directions for future research in Chapter~\ref{6:}.

%% file: text/C1/C1.tex
\chapter{\label{2:}OTOCs}
An `out-of-time ordered correlation function' (OTOC) is a widely used measure of information scrambling in quantum systems  \cite{garcia-mataOutoftimeorderCorrelationsQuantum2023}. This chapter begins with an examination of the Lyapunov exponent and its connection to classical chaos. We then discuss the relationship between chaos and scrambling, extending analysis to quantum systems. Following this, we introduce `ring polymer molecular dynamics' (RPMD), by which semiclassical OTOCs may be calculated. 
\section{\label{sec:classicalchaos}Chaos in classical systems}
In the study of dynamical systems, the Lyapunov exponent $\lambda$ is used to quantify the sensitivity of a system property with respect to slight changes in initial conditions. Consider a one-dimensional classical system with Hamiltonian
\begin{equation}
\label{eqn:1dclassicalhamiltonian}
    H(p,q) = T(p) + V(q) \text{,}
\end{equation}
 where the kinetic energy $T(p)=\frac{p^2}{2m}$, $V(q)$ is the potential energy and $p$ and $q$ are the momentum and position respectively. If a trajectory is chaotic,
 \begin{equation}
 \label{eqn:lyapunovexpnt}
     |\delta q_t|\approx |\delta q|e^{\lambda t}\text{,}
 \end{equation}
where $\delta q$ is an initial perturbation in $q$ and $\delta q_t$ is the corresponding change in $q$ at time $t$. A system is commonly\cite{goldsteinClassicalMechanics2002} considered to be chaotic if the difference between two typical (i.e. not periodic orbits or stationary points) trajectories grows exponentially with time, indicated by a positive exponent in Eq.\eqref{eqn:lyapunovexpnt}.
\\[5pt]
\indent 
A more general metric of sensitivity to initial perturbations is the `stability matrix'
\begin{equation}
\label{eqn:Mclass}
    \mt{M}(t) \equiv \begin{pmatrix}
        M_{pp} & M_{pq} \\
        M_{qp} & M_{qq}
    \end{pmatrix}
    = \begin{pmatrix}
       {\partial p_t}/{\partial p} & {\partial p_t}/{\partial q} \\
        {\partial q_t}/{\partial p} & {\partial q_t}/{\partial q}
    \end{pmatrix} \text{.}
\end{equation}
Each element of $\mt{M}$ gives the sensitivity of the system phase space variables $x\in q,p$ at time $t$ with respect to in infinitesimal perturbation at time $0$. Note here that for $\delta q\to 0$, exponential growth rate of $M_{qq}$ of Eq.\eqref{eqn:Mclass} corresponds to the Lyapunov exponent in Eq.\eqref{eqn:lyapunovexpnt}. 
\\[5pt]
\indent By differentiating the equations of motion for $p_t$ and $q_t$ with respect to $p$ and $q$ we obtain
\begin{equation}
\label{eqn:Meom}
    \frac{\d}{\d t} \mt{M} = \begin{pmatrix}
        0 & -V''(q_t) \\
        m^{-1} & 0
    \end{pmatrix}
    \mt M\text{.}
\end{equation}
The potential energy curvature $V''(q) \equiv \frac{\d^2 V}{\d q^2}$ generalises to the Hessian in the case of a system of multiple dimensions.  When trajectories pass through regions for which $V''(q)<0$, Eq.\eqref{eqn:Meom}  shows that the elements of the stability matrix grow exponentially, giving a positive Lyapunov exponent.  For this reason even non-chaotic  systems, containing areas of negative potential curvature, may still exhibit globally positive Lyapunov exponents.
\\[5pt]
\indent 
In this report, we focus mainly on non-classically chaotic `barrier scrambling' models. The simplest example of this is a double well, where only the trajectories in the local vicinity of the barrier are chaotic. 
\section{\label{sec:factor2}Thermally-averaged classical scrambling}
We denote thermal expectation values at inverse temperature $\beta$ as $\left<\cdot \right>_{\rm cl}$. For the one-dimensional system of Sec.\,\ref{sec:classicalchaos}, 
\begin{equation*}
    \left<\cdot\right>_{\rm cl} \equiv \frac{1}{hZ_{\rm cl}}\int \d p\,\d q\: \cdot e^{-\beta H(p,q)} \: \text{  ,}
\end{equation*}
where $h$ is Plank's constant, and the classical partition function
\begin{equation*}
    Z_{\rm cl} = \frac{1}{h}\int \d p\,\d q\: e^{-\beta H(p,q)} \text{.}
\end{equation*}
The integration limits are infinite in general unless stated otherwise throughout this thesis, with multiple integral signs abbreviated to a single one when possible. 
\\[5pt]
\indent
To obtain a thermally-averaged Lyapunov exponent, one could construct
\begin{equation}
\label{eqn:badclOTOC}
\begin{split}
    C_1(t) = \left<\left(\frac{\partial q_t}{\partial q}\right) \right>_{\rm cl}
\text{,}
\end{split}
\end{equation}
then measure the exponential growth rate of the correlation function. Following integration by parts though, Eq.\eqref{eqn:badclOTOC} gives
\begin{equation}
        C_1(t)= \beta \left< V'(q)\: q_t \right>_{\rm cl} \text{.}
\end{equation}
This is the form of a two-time correlation function, which due to the Cauchy-Schwarz rule of weighted inner-products is bounded such that $\forall t$
\begin{equation}
    \left|\left< V'(q)\:q_t \right>_{\rm cl}\right| \leq \sqrt{\left< V'(q)^2\right>_{\rm cl}\left<q^2\right>_{\rm cl}} \text{.}
\end{equation}
This bounding nullifies any possibility of long-term exponential growth, as also observed in other normal two-time correlation functions. For this reason it is more useful to consider, 
\begin{equation}
    \label{eqn:cl1dOTOC}
\begin{split}
    C_{\rm cl}(t) &= \hbar^2\left<\left(\frac{\partial q_t}{\partial q}\right) ^2\right>_{\rm cl} \text{,}
\end{split}
\end{equation}
which can no longer be rearranged into a multi-time correlation function. The presence of $\hbar^2$ in Eq.\eqref{eqn:cl1dOTOC} is due to quantum-classical correspondence as will be explained in the following section. We extract the classical thermal Lyapunov exponent $\lambda_{\rm cl}$ as
\begin{equation}
    \lambda_{\rm cl} := \frac{\d}{\d t}\ln C_{\rm cl}(t)\text{,} 
\end{equation}
where $t\in I$, the time interval within which $C_{\rm cl}(t)$ grows exponentially. Here, $\lambda$ is assumed to be time-independent, but this may not always be true. In practice, the Lyapunov exponent is calculated as an average of the early-time exponential growth of $C_{\rm cl}(t)$ as calculation at later times becomes exponentially more computationally expensive.
\\[5pt]
\indent 
At late times, we can calculate the growth rate of $C_{\rm cl}(t)$ for barrier scrambling systems by considering linearised dynamics about fixed points in phase space, following the approach of Xu \etal\!\!\cite{xuDoesScramblingEqual2020}. As mentioned in Sec.~\ref{sec:classicalchaos}, we consider a one-dimensional system with a single saddle point in $V(q)$ which we call the `barrier'. We position the barrier at $q=0$ and give  curvature $\sqrt{\frac{1}{m}V''(0)} =i\omega_{\rm b}$.
\\[5pt]
\indent 
Due to the form of Eq.\eqref{eqn:Meom}, at late times the growth of Eq.\eqref{eqn:cl1dOTOC} is dominated by trajectories that spend all time in the vicinity of the barrier. In this region of configuration space, the dynamics corresponding to the exact Hamiltonian $H(p,q)$ can be generated by an approximate Hamiltonian
\begin{equation}
\label{eqn:approximateclassicalham}
    \begin{split}
    H(p,q) \approx H_\delta(p,q) &:= \frac{p^2}{2m} - \frac{1}{2}m\omega_{\rm b}^2q^2\text{.}
    \quad\quad\left|q\right|<\delta\ll1
    \end{split}
\end{equation}
Applying the transformation 
\begin{equation}
    \begin{pmatrix}
        a^+\\
        a^-
    \end{pmatrix}
    = \frac{1}{\sqrt{2m\omega_{\rm b}}}
    \begin{pmatrix}
        1 & m\omega_{\rm b} \\
        1 & -m\omega_{\rm b} 
    \end{pmatrix}
    \begin{pmatrix}
        p \\ q
    \end{pmatrix}
    \text{,}
\end{equation}
we use Eq. \eqref{eqn:approximateclassicalham} to generate decoupled equations of motion for these new variables as
\begin{equation}
    a^\pm_t = a^\pm\:e^{\pm\omega_{\rm b}t} \text{.}
\end{equation}
We now define the constraint function
\begin{equation}
    \label{eqn:constrainfunction}
S(x) = \left\{\begin{array}{ll} 0 & \quad\quad |x|>\delta
 \\1 & \quad\quad |x|\leq\delta
\end{array}   \right. 
\end{equation}
 with which we constrain $q$ and $q_t$ (which can now be written as functions of $a^\pm$ and $a^\pm_t$ respectively) within the thin strip of phase space surrounding the saddle point. We therefore write $C_{\rm cl}$ at late times in terms of trajectories that have spent all time in the vicinity of the barrier as
\begin{equation}
\begin{split}
        C_{\rm cl}(t\to \infty) &= \hbar^2\int \d a^+\d a^- e^{-\beta H_\delta}\left(\frac{\partial q_t}{\partial q}\right)^2 S(a^+-a^-)S(a^+e^{\omega_{\rm b} t} - a^-e^{-\omega_{\rm b}t}) \text{.}
    \end{split}
\end{equation}
At large $t$, the product of constraint functions is non-zero for $|a^+|<\delta\exp(-\omega_b t)$ and $|a^-|<\delta\exp(-\omega_bt)+\delta\simeq\delta$. Also, using the equations of motion of $a^{\pm}$, 
\begin{equation}
\begin{split}
   \lim_{t\to\infty} \left[\frac{\partial q_t}{\partial q}\right] &= e^{\omega_{\rm b}t}\text{,}\quad\quad\left|q\right|<\delta\ll1
\end{split}
\end{equation}
such that  
\begin{equation}
    \begin{split}
   C_{\rm cl}(t\to \infty)  
    &= \hbar^2e^{2\omega_{\rm b} t}\int_{-\delta e^{-\omega_{\rm b}t}}^{\delta e^{-\omega_{\rm b}t}} \d a^+\int_{-\delta}^\delta \d a^- \:e^{-\beta H_\delta(a^+,a^-)}
    \\
    &\sim e^{\omega_{\rm b} t} \text{.}
\end{split}
\end{equation}
The exponentially shrinking limits on $\d a^+$ reduce the growth of $C_{\rm cl}(t)$ at late times by a factor of two. We conclude that 
\begin{equation}
    \lambda_{\rm cl} \approx \omega_{\rm b}\text{,}
\end{equation}
for one-dimensional classical barrier scrambling systems at late times. We therefore can verify that $C_{\rm cl}$  for barrier scrambling grows exponentially at late times and following similar arguments, see that $C_1$ will not.
\section{Quantum scrambling}
It is well known that  commutators between operators $\hat W$ and $\hat V$ in quantum mechanics correspond to Poisson brackets \cite{diracFundamentalEquationsQuantum1997} between variables $w$ and $v$ in the $\hbar\to0$ limit;
\begin{equation}
    \lim_{\hbar\to0}\frac{1}{i\hbar}[\hat W,\hat V] = \{w,v\}.
\end{equation}
Therefore, for $\hat W = \hat q$ and $\hat V = \hat p$, Eq.\eqref{eqn:cl1dOTOC} is the classical limit of
    \begin{equation}
    \label{eqn:stdregOTOC}
\begin{split}
    C_{\rm qm}(t) &= \left<[\hat q_t,\hat p] ^\dagger[\hat q_t,\hat p]\right>_{\rm std} \text{.}
\end{split}
\end{equation}
 When possible we will abbreviate $C_{\rm qm}(t)$ as $\left<[\hat q_t,\hat p]^2\right>_{\rm std}$  for legibility, where the absolute square of the commutator is implied.  The $t$-subscripts indicate that operators have been evolved in time up to $t$ (i.e. $\hat q_t \equiv e^{\frac{it}{\hbar}\hat H}\hat q\: e^{-\frac{it}{\hbar}\hat H}$). For thermal expectation values, we now use the quantum thermal average 
\begin{equation}
    \left<\cdot\right>_{\rm std}\equiv \Tr{\hat\rho\:\cdot}\text{,}
\end{equation}
for which $\hat\rho=Z_{\rm qm}^{-1}e^{-\beta\hat H}$ is the thermal density matrix and the quantum canonical partition function $Z_{\rm qm} = \Tr{e^{-\beta \hat H}}$. 
\\[5pt]
\indent
The thermal density matrix can be spread between operators in the expression of Eq.\eqref{eqn:stdregOTOC}. This spreading, referred to as `regularisation', does not alter the classical limit. One commonly \cite{chowdhuryOnsetManybodyChaos2017} used regularisation of  $C_{\rm qm}$ is the `symmetrised' version for which pairs of operators are separated by $\hat{y}^2 = \hat\rho^{1/2}$.
\begin{equation}
 \left<[\hat q_t,\hat p]^2\right>_{\rm sym} := \Tr{\hat y^2[\hat q_t,\hat p]^\dagger\hat y^2 [\hat q_t,\hat p] } \text{.}
\end{equation}
Other forms of regularisation are possible, and will be introduced when appropriate.
\section{A bound on quantum scrambling}
Quantum OTOCs have been extensively studied the field of many-body physics. In this context, there are two main timescales relevant to the growth of OTOCs. The first is the dissipation time $t_{\rm d}$ which gives the timescale on which the memory of local interactions is lost in the system. This timescale is commonly related to the decorrelation time of two-point correlators. The second is the scrambling time $t_{\rm s}$, the timescale for which any initial information (in the form of wavefunction localisation) fed into the system is irretrievably spread. 
\\[5pt]
\indent 
For systems in which there is a hierarchy in timescales such that $t_{\rm d} \ll t_{\rm s}$, some display OTOCs which grow exponentially, analogously to classical mechanics,
\begin{equation}
    \left<[\hat W_t,\hat V] ^2\right> \sim e^{\lambda_{\rm qm} t}\text{.} \quad\quad\quad\quad t_{\rm d} < t \ll t_{\rm s} 
\end{equation}
Here, $\lambda_{\rm qm}$ is commonly referred to as the `quantum Lyapunov exponent'.
\\[5pt]
\indent
In 2016, Maldacena \etal conjectured a universal (system independent) upper bound on this exponent \cite{maldacenaBoundChaos2016} 
\begin{equation}
\label{eqn:bound}
    \lambda_{\rm qm} \leq \frac{2\pi}{\beta\hbar} \text{,}
\end{equation}
which we refer to as the `Maldacena bound'. Due to the reciprocal dependence on $\hbar$ in Eq.\eqref{eqn:bound}, all classical systems trivially satisfy this bound. 
Following this, Tsuji \etal re-derived this bound\cite{tsujiBoundExponentialGrowth2018} using the assumption that exponential growth of the OTOC was independent of regularisation.
\\[5pt]
\indent
Although both proofs require relatively stringent criteria and assumptions to be valid, numerical studies have all so far shown that both quantum many-body \cite{kobrinManyBodyChaosSachdevYeKitaev2021,bohrdtScramblingThermalizationDiffusive2017,patelQuantumChaosCritical2017,banerjeeSolvableModelDynamical2017} and single-body systems\cite{hashimotoBoundEnergyDependence2022,hashimotoOutoftimeorderCorrelatorsQuantum2017}, including barrier scrambling systems \cite{hashimotoExponentialGrowthOutoftimeorder2020}, satisfy the bound. This consistency suggests that, even though the Maldacena bound was introduced in the context of many-body physics, it may be a fundamental property of all quantum systems. 
\section{\label{sec:RPMDintro}RPMD approach to OTOCS}
Ring polymer molecular dynamics (RPMD)\cite{craigChemicalReactionRates2005} is a method introduced by Manolopoulos \etal using the fictitious dynamics of `ring polymers' to approximate `Kubo-regularised' time correlation functions (TCFs). These dynamics \textit{have the decency}\footnotemark{}\footnotetext{W.H. Miller} to satisfy quantum detailed-balance. 
Following the introduction of RPMD it was pointed out by Althorpe \etal that it is an approximation to Matsubara dynamics \cite{heleBoltzmannconservingClassicalDynamics2015}, a theory by which Fourier smoothing of the path integral gives quantum dynamics without real-time coherence. In this section we introduce the basic concepts behind RPMD, and state the form of an RPMD OTOC. For a detailed derivation see Ref.\cite{sadhasivamDynamicalSignaturesInstantons2023}.
\\[5pt]
\indent
In our discussions here, we continue to use the Hamiltonian of Eq.\eqref{eqn:1dclassicalhamiltonian}, with hats on corresponding quantum operators. Generalisation to multiple dimensions is straightforward. Due to Feynman's path integral formulation of quantum statistical mechanics\cite{feynmanQuantumMechanicsPath2010} the quantum partition function, at inverse temperature $\beta$
\begin{equation}
    Z_{\rm qm} = \lim_{N\to\infty} Z_{N} \text{,}
\end{equation}
where
\begin{equation}
\label{eqn:discretisedpathintegralZ}
    Z_N = \frac{1}{(2\pi\hbar)^{N}}\int \d\vc{p}\:\d\vc{q} \:e^{-\beta_N H_{N} (\vc{p},\vc{q})}\text{,}
\end{equation}
with $\beta_N = \beta/N$. Equation~\eqref{eqn:discretisedpathintegralZ} is an imaginary time path integral, discretized into $N$ imaginary-time slices, each of which is given a set of phase space variables \{$p_n,q_n\}$. We abbreviate integrals over multiple coordinates (and will continue to do so in later sections) as $\d\vc{x}\equiv\d x_1\d x_2...\d x_N$. The ring polymer Hamiltonian,
\begin{equation}
    H_{N}(\vc{p},\vc{q}) =\underbrace{ \sum_{n=1}^{N}\left[\frac{p_n^2}{2m} + V(q_n)\right]}_{\mathlarger{{T_{N}(\vc{p}) + V_N(\vc{q}})}} + S_{N}(\vc{q})
\end{equation}
is a sum over a ring of identical classical systems with Hamiltonian $H(p_n,q_n)$ with coordinates adjacent in imaginary time connected by the spring potential
\begin{equation}
\label{eqn:rpmdSprings}
    S_N(\vc{q}) = \frac{m}{2(\beta_N\hbar)^2}\sum_{n=1}^N (q_n-q_{n+1})^2\text{.}\quad\quad (q_{N+1} \equiv q_1)
\end{equation}
In RPMD, the quantum partition function $Z_{\rm qm}$ is approximated by $Z_N$ for some finite $N$. The quantum statistics of RPMD are exact in the $N\to\infty$ limit (where the discrete paths $\vc{q}$ become functions of imaginary time $q(\tau)$ of period $\beta\hbar$), but in practice $N$ is treated as a convergence parameter, commonly ranging from $16$-$128$\cite{habershonRingpolymerMolecularDynamics2013}. The ring polymer Hamiltonian, $H_N$, is used to generate artificial, classical dynamics - `ring polymer molecular dynamics'. In other words, for some system property $f(\vc{p},\vc{q})$ time evolution is given by
\begin{equation}
    \frac{\d}{\d t}f(\vc{p},\vc{q}) = \mathcal{L}_{\rm rp}f(\vc{p},\vc{q}) \text{,}
\end{equation}
where $\mathcal{L}_{\rm rp}\cdot\equiv\left\{H_N(\vc{p},\vc{q}),\cdot\right\}$ and $\{\cdot,\cdot\}$ is the Poisson bracket corresponding to the $2N$-dimensional ring polymer phase space.
\\[5pt]
\indent 
The RPMD approximation to Kubo-regularised correlation functions can be shown to be exact in the high-temperature, harmonic and $t\to0$ limits \cite{habershonRingpolymerMolecularDynamics2013}. In addition to these pleasant properties, RPMD preserves the quantum Boltzmann distribution, known as `detailed balance'. To demonstrate this we consider some RPMD TCF between operators $A$ and $B$,
\begin{equation}
\begin{split}
\label{eqn:detailedbalancederivation}
    C_{\rm AB}^{\rm rp} :=& \:\frac{1}{(2\pi\hbar)^NZ_N} \int \d\vc{p}\:\d\vc{q}\:e^{-\beta_NH_N(\vc{p},\vc{q})}A(\vc{p},\vc{q})\:e^{\mathcal{L}_{\rm rp}t}B(\vc{p},\vc{q})
    \\
    =&\: \frac{1}{(2\pi\hbar)^NZ_N} \int \d\vc{p}\:\d\vc{q}\:e^{-\beta_NH_N(\vc{p},\vc{q})}B(\vc{p},\vc{q})\:e^{-\mathcal{L}_{\rm rp}t}A(\vc{p},\vc{q}) \text{,}
\end{split}
\end{equation}
where the second line in Eq.\eqref{eqn:detailedbalancederivation} is obtained by expanding $e^{\mathcal{L}_{\rm rp}t}$ as a Taylor series in $t$, then applying integration by parts to move all derivatives applied on $B(\vc{p},\vc{q})$ onto $e^{-\beta H}A(\vc{p},\vc{q})$ and rearranging. The manipulation of Eq.\eqref{eqn:detailedbalancederivation} implies detailed balance,
\begin{equation}
    \label{eqn:detailedbalance}
    C_{AB}^{\rm rp}(t)=C_{BA}^{\rm rp}(-t)\text{.}
\end{equation}
By choosing $A\!\!=\!\!1$, detailed balance in RPMD TCFs implies
\begin{equation}
    C_{1B}^{\rm rp}=C_{B1}^{\rm rp} \equiv\left\langle\hat B\right\rangle_{\rm std} \text{,}
\end{equation}
as $C_{B1}$ is simply the path integral formulation of the exact quantum thermal expectation value of $\hat B$.
\\[5pt]
\indent
Following Sadhasivam \cite{sadhasivamDynamicalSignaturesInstantons2023}, we take the Kubo regularisation of a quantum OTOC as
\begin{equation}
\label{eqn:quantumkuboOTOC}
    \left<[\hat q_t,\hat p]^2\right>_{\rm kubo} := \frac{1}{\beta}\int_{0}^\beta \d\lambda\: \Tr{\hat \rho^{\:-(\beta-\lambda)}[\hat q_t,\hat p]^\dagger \hat\rho^{\:-\lambda}[\hat q_t,\hat p]}\text{,}
\end{equation}
in which the density matrix is `smeared' between each commutator. This regularised form of correlation functions in general ensures that the TCF is real and time-reversible. The RPMD approximation of Eq.\eqref{eqn:quantumkuboOTOC} is
\begin{equation}
\label{eqn:rpOTOC}
    \left<\left(\frac{\partial X_{\!0\,t}}{\partial X_{\!0}}\right)^2\right>_{\rm rp} := \frac{1}{(2\pi\hbar)^{N}Z_N}\int \d\vc{p}\:\d\vc{q} \:e^{-\beta_N H_{N} (\vc{p},\vc{q})}\left(\frac{\partial X_{\!0\,t}}{\partial X_{\!0}}\right)^2 \text{,}
\end{equation}
wherein the centroid coordinate
\begin{equation}
\label{eqn:rpmdCentroid}
    X_{\!0} = \frac{1}{N}\sum_{n=1}^N q_n \text{,}
\end{equation}
the arithmetic mean of the positions of each identical classical system. Note that the form of Eq.\eqref{eqn:rpmdCentroid} is exactly the same as a multi-dimensional classical OTOC in the $2N$-dimensional phase space defined by $H_N$.
By considering Taylor expansions about $t=0$, it can be shown that the RPMD agrees with the Kubo-regularised quantum OTOC up to $O(t^6)$. 
\\[5pt]\indent 
As they obey quantum detailed balance, RPMD and Matsubara dynamics conserve the quantum Boltzmann distribution \cite{willattMatsubaraDynamicsIts2017}, in contrast to other common semiclassical methods such the family based on Initial Value Representation \cite{thossSemiclassicalDescriptionMolecular2004}. Ref.\cite{sadhasivamInstantonsQuantumBound2023} argues that the Maldecena bound is enforced due to conservation of these statistics. We will further explore and test this claim in the following chapter. 

%% file: text/C2/C2.tex
\chapter{\label{3:}RPMD OTOCs} 
Sadhasivam \etal have shown that formation of `instantons' in symmetric barrier scrambling systems reduces the exponential growth rate of RPMD OTOCs\cite{sadhasivamInstantonsQuantumBound2023} at low temperatures, such that they satisfy the Maldacena bound. In this chapter, we first outline their proof, then examine specific examples where it may break down, exploring the reasons for these exceptions. 
\section{Multidimensional classical scrambling\label{sec:multidimensionalClassicalScrambling}}
In this section we generalise the result of Sec.\,\ref{sec:factor2} to a single particle classical system of $N$ dimensions. We will use this result in the next section for the $N$-dimensional system defined by RPMD. We now define the stability matrix
\begin{equation}
    \mt{M}(t) := \frac{\partial\{\vc{p}_t,\vc{q}_t\}}{\partial\{\vc{p},\vc{q}\}}
\end{equation}
where the phase space variables $\vc{p}$ and ${\vc{q}}$ are now $N$-dimensional and our Hamiltonian
\begin{equation}
    H(\vc{p},\vc{q}) = \frac{\vc{p}^\top\vc{p}}{2m} + V(\vc{q})\text{.}
\end{equation}
We obtain the equations of motion for the stability matrix as
\begin{equation}
    \frac{\d}{\d t}\mt{M}(t) = \mt{A}(t)\mt{M}(t)  \text{,} \label{eqn:generalclassicalmonodromy}
\end{equation}
where \begin{equation}
\label{eqn:classicalgeneralA}
   \mt{A}(t) = \begin{pmatrix}
        \mt{0}&-\mt{V''}(\vc{q}_t)\\
        m^{-1}\mt{1}&\mt{0}\\
    \end{pmatrix}\text{.}
\end{equation}
The Hessian at time $t$, denoted as $\mt{V''}(\vc{q}_t)\equiv\frac{\partial^2 V(\vc{q}_t)}{\partial \vc{q}_t\partial \vc{q}_t}$. We use $\mt{0}$ and $\mt{1}$ as the $N\times N$  zero and identity matrices respectively. 
\\[5pt]
\indent
For a fixed point in dynamics at $\{\vc{p}_t,\vc{q}_t\} = \{\vc{\tilde p},\vc{\tilde q}\}\:\forall t$ meaning that the stability matrix propagator $\mt{A}(t)\to\mt{\tilde A}$, a constant. The equation of motion of the stability matrix at this point, $\mt{\tilde{M}}$, is therefore governed by a first order linear differential equation. This makes each element of $\mt{\tilde{M}}$ a linear combination of exponential terms, with growth rates determined by the eigenvalues $\lambda_i,i\in\{\pm1,\pm2\cdots\pm N\}$ of $\mt{\tilde A}$. 
\begin{equation}
\begin{split}
\mt{\tilde{M}}_{ij}(t) &= \sum_{i} c_{ij}\:e^{\lambda_it}\text{,}
\end{split}
\end{equation}
which means that in the late-time limit,
\begin{equation}
\label{eqn:latetimeOTOC-classicalN}
\begin{split}
\mt{\tilde{M}}_{ij}(t\to\infty) &\sim e^{\lambda_{\rm max}t}\text{,}
\end{split}
\end{equation}
provided that $c_{ij}\neq0$ for $i$ corresponding to $\lambda_i=\lambda_{\rm max}$,  the eigenvalue of $\mt{\tilde A}$ with largest positive real part. In practice it is found that $c_{ij}=0$ is only ever true in special cases where a prohibitive symmetry constraint exists, which is not encountered in this report. 
\\[5pt]
\indent 
As $\mt{\tilde A}$ is block antidiagonal, its eigenvalues satisfy \cite{silvesterDeterminantsBlockMatrices2000}
\begin{equation}
\label{eqn:relationofHesstoAClassical}
    \det{\lambda_i^2\mt{1}+\frac{1}{m}\mt{V''}(\vc{\tilde q})} = 0
\end{equation}
meaning that we may relate these eigenvalues to those of the mass-weighted Hessian eigenvalues as
\begin{align}
    \lambda_{\pm i} = \pm \eta_i \text{,} \label{eqn:relationoflambdatoeta}
\end{align}
where
\begin{equation}
    \eta_i :=\sqrt{-\frac{1}{m}\frac{\partial^2V(\vc{\tilde q})}{\partial\xi_i^2}}
\end{equation}
and $\xi_i ,i\in\{1,2\cdots N\}$ are the normal modes of the Hessian. 
We therefore conclude that the max growth rate of the stability matrix at fixed points in multi-dimensional systems is equal to the most negative eigenfrequency of the mass-weighted Hessian at that point.
\\[5pt]
\indent 
Following the same logic as in Sec.\,\ref{sec:factor2} we now consider the dynamics about the fixed point. We use the constraint function from Eq.\eqref{eqn:constrainfunction} to define the `survival probability'
\begin{equation}
    P(t) =\int\d|\vc{z}| \prod_{i=1}^{2N}S(|z_{i}|)S(|z_{i\:t}|)\text{,}
\end{equation}
which indicates the probability that a trajectory, starting in a region of phase space where the stability matrix grows at the rate defined by $\mt{\tilde{A}}$, remains in that region at time $t$. Here, $\{\vc{z}\}$ is some complex coordinate system made from $\{\vc{p},\vc{q}\}$ corresponding to the eigenvectors of $\mt{\tilde{A}}$ centered at $\{\vc{\tilde{p}},\vc{\tilde{q}}\}$. As we know the dynamics about this fixed point in these coordinates,
\begin{equation}
\label{eqn:depopulationFactor-Nclassical}
\begin{split}
    P(t) &=\int\d|\vc{z}| \prod_{i=1}^{2N}S(|z_{i}|)S(|z_{i}\:e^{\lambda_i t}|)
       \\ &\sim \prod_{\Re{\lambda_i}>0}e^{-\lambda_it}\text{.}
\end{split}
\end{equation}
\\
\indent
Combining Eq.\eqref{eqn:latetimeOTOC-classicalN} with Eq.\eqref{eqn:depopulationFactor-Nclassical} gives the late-time growth rate of a barrier scrambling OTOC in an $N$-dimensional space as
\begin{equation}
\begin{split}
    C(t\to\infty) &\sim e^{2\lambda_{\rm max}t}\prod_{\Re{\lambda_i}>0}e^{-\lambda_it}\text{,}
\end{split}
\end{equation}
a result analogous to that obtained by Xu \etal in \cite{xuDoesScramblingEqual2020}. This result can be written in terms of the normal mode frequencies using Eq.\eqref{eqn:relationoflambdatoeta} to give our final result
\begin{equation}
\begin{split}
\label{eqn:latetimeNdimOTOC}
    C(t\to\infty) &\sim e^{2\eta_{\rm max}t}\prod_{\eta_i>0}e^{-\eta_it}\text{.}
\end{split}
\end{equation}
 Note that for the case of a first order saddle in potential, the growth rate of $C(t\to\infty)$ simplifies to $\eta_{\rm max}$, the negative eigenfrequency of the mass-weighted Hessian.
\section{Instantons and the bound on chaos}
\begin{figure}[hbt!]
\centering
\scalebox{1}{\includegraphics[]{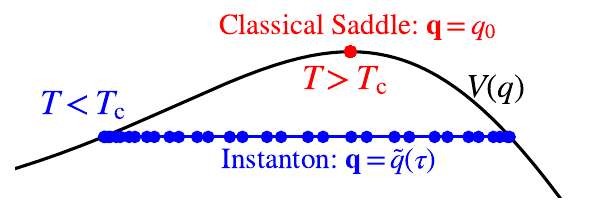}
}
\caption[Instanton formation for $T<T_{\rm c}$]{Instanton formation for  $T<T_{\rm c}$. $V(q)$ is the asymmetric Eckart model used in Ref.\cite{richardsonRingpolymerMolecularDynamics2009}.}
\label{fig:inst_pot}
\end{figure}

RPMD is classical mechanics in $2N$-dimensional phase space, for which the new effective `ring polymer potential'
\begin{equation}
    U_N(\vc{q}) = V_N(\vc{q}) + S_N(\vc{q})\text{,}
\end{equation}
using the definitions established in Sec.\,\ref{sec:RPMDintro}. Due to the presence of $\beta_N$ in the potential (in the springs between adjacent beads), the potential energy landscape defined by $U_N$ is temperature dependent.
\\[5pt]
\indent
For barrier scrambling systems, above the `crossover temperature'
\begin{equation}
    T_{\rm c} = \frac{\hbar\omega_b}{2\pi k_B} \text{,}
\end{equation}
there is a first order saddle in $U_N$ corresponding to all of the beads located at the top of the barrier (see the red configuration in Fig.~\ref{fig:inst_pot}). We refer to this configuration as the `classical saddle', for which we may use Eq.\eqref{eqn:latetimeNdimOTOC} to calculate the late-time growth of an RPMD OTOC as
\begin{equation}
    \lambda_{\rm rp} \approx \omega_b\text{.} \quad\quad\quad\quad T>T_c
\end{equation}
Due to the form of  Eq.\eqref{eqn:rpmdSprings}, as $T$ is decreased below $T_c$, the springs holding adjacent beads become so weak that a delocalised structure now becomes the first order saddle in $U_N$, with a lower potential energy (see the blue configuration in Fig.~\ref{fig:inst_pot}). This structure is known as the `instanton', which we in general denote with a tilde (i.e. $\vc{\tilde{q}}$ corresponds to $\vc{q}$ valued at the instanton).
\\[5pt]
\indent
The proof of Sadhasivam begins with the observation that trajectories that pass over the barrier below $T_c$ fluctuate about the instanton. Hence, using Eq.\eqref{eqn:latetimeNdimOTOC} the growth of the OTOC 
 \begin{equation}
       \lambda_{\rm rp} \approx \eta_0\text{,} \quad\quad\quad\quad T<T_c
 \end{equation}
 where
 \begin{equation}
     \eta_0 = \sqrt{-\frac{1}{m}\frac{\partial^2U_N(\vc{\tilde q})}{d\xi_0^2}} \text{,}
 \end{equation}
 and $\xi_0$ is the unstable normal mode of the RPMD Hessian at the instanton.
In the case of a symmetric barrier, $\xi_0$ is primarily directed along the centroid coordinate $X_0$ (defined in Eq.\eqref{eqn:rpmdCentroid}). For this reason they approximate
\begin{equation}
\begin{split}
\label{eqn:etaapproximation}
    \eta_0 &\approx \sqrt{-\frac{1}{m}\frac{\partial^2U_{N}(\vc{\tilde q})}{\partial X_{\!0}^2}}
    \\
    &= \sqrt{-\frac{1}{m}\frac{\partial^2V_{N}(\vc{\tilde q})}{\partial X_{\!0}^2}}.
\end{split}
\end{equation}
The second equality in Eq.\eqref{eqn:etaapproximation} follows due to there being no coupling between springs in $S_N$ and $X_0$ in Eq.\eqref{eqn:rpmdSprings}.
Now as we know the instanton is a first order saddle in $U_N$, any displacements perpendicular to $X_0$ must be stable with respect to perturbations, that is 
\begin{equation}
\label{eqn:Un_stabletoperturbations}
    \frac{\partial^2 U_N(\vc{\tilde q})}{\partial {X_{k}}^2 }\geq 0
    \quad\quad k = \pm1, \pm2 \cdots
\end{equation}
where the orthogonal basis vectors
\begin{equation}
\label{eqn:discretefouriertrs}
    {X_{k}} = \frac{1}{\sqrt{N}}\sum_{n=1}^NT_{kn}q_n\text{,}
\end{equation}
and the $N\times N$ discrete Fourier transform matrix
\begin{equation}
\label{eqn:discretefourierMatrix}
    {T_{kn}} = 
    \begin{cases}
        \sqrt{\frac{2}{N}}\cos\left(\frac{2kn\pi}{N}\right) &k =-1,-2,-3\cdots \\
        \sqrt{\frac{1}{N}}& k = 0\\
              \sqrt{\frac{2}{N}}\sin\left(\frac{2kn\pi}{N}\right)& k = 1,2,3\cdots
    \end{cases} 
\end{equation}
These coordinates are eigenvectors of the Hessian corresponding to $S_N$ with $N\to\infty$ eigenvalues $\left(\frac{2k\pi}{\beta\hbar}\right)^2\!\!\!$ . Due to $U_N=V_N+S_N$, Eq.\eqref{eqn:Un_stabletoperturbations} implies
\begin{equation}
\label{eqn:boundingoffluctuations}
    \frac{1}{m}\frac{\partial^2 V_N(\vc{\tilde q})}{\partial {X_{k}}^2 }\geq -\left(\frac{2k\pi}{\beta\hbar}\right)^2
    \quad\quad k = \pm1, \pm2 \cdots
\end{equation}
Now it can be shown using Eq.\eqref{eqn:discretefouriertrs} that 
\begin{equation}
\label{eqn:cent-fluctuationsderivativerelation}
    \frac{\partial^2 V_N(\vc{q})}{\partial X_0^2} = \frac{1}{2}\left[\frac{\partial^2 V_N(\vc{q})}{\partial X_k^2}+ \frac{\partial^2 V_N(\vc{q})}{\partial X_{-k}^2}\right]
    \quad\quad k = \pm1, \pm2 \cdots
\end{equation}
Combining Eq.\eqref{eqn:boundingoffluctuations} with Eq.\eqref{eqn:cent-fluctuationsderivativerelation} we have
\begin{equation}
 -\frac{1}{m}\frac{\partial^2 V_N(\vc{q})}{\partial X_0^2} \leq \left(\frac{2\pi}{\beta\hbar}\right)^2 \text{,}
\end{equation}
which, given the assumption of Eq.\eqref{eqn:etaapproximation} therefore gives
\begin{equation}
\eta_0 \leq \frac{2\pi}{\beta\hbar} \text{,}
\end{equation}
making the growth of RPMD OTOCs satisfy the Maldacena bound.
\section{Breaking the bound}
In this section we illustrate two examples for which RPMD OTOCs violate the Maldacena bound. The first is a system for which instantons do not form, due to there being no saddle in potential. The second example is where the assumption of Eq.\eqref{eqn:etaapproximation} breaks down.
\subsection{Systems without instantons\label{sec:ChaoticSystems}}
\begin{figure}[]
\centering
\scalebox{1}{\includegraphics[trim={1cm 0.5cm 0.4cm 1cm}]{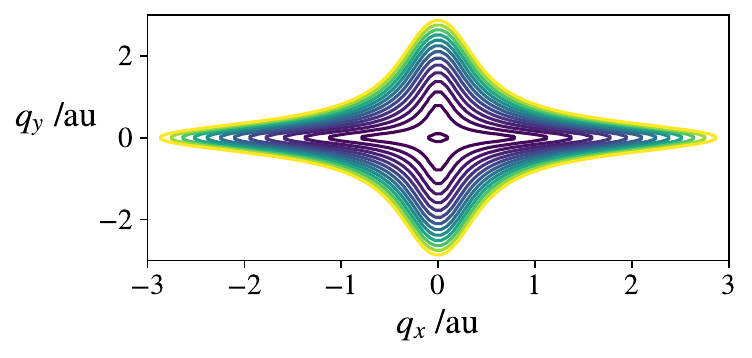}
}
\caption[A plot of the two-dimensional classical potential]{A contour plot of $V(q)$ defined in Eq.\eqref{eqn:2dchaospotential} for the values of $a=0.1$, $b=10$ used in simulation. Further details regarding simulation and convergence parameters can be found in Appendix \ref{app:parametersforCounterExample}. }
\label{fig:2dqopot}
\end{figure}

\begin{figure}[H]
\centering
\scalebox{1}{\includegraphics[]{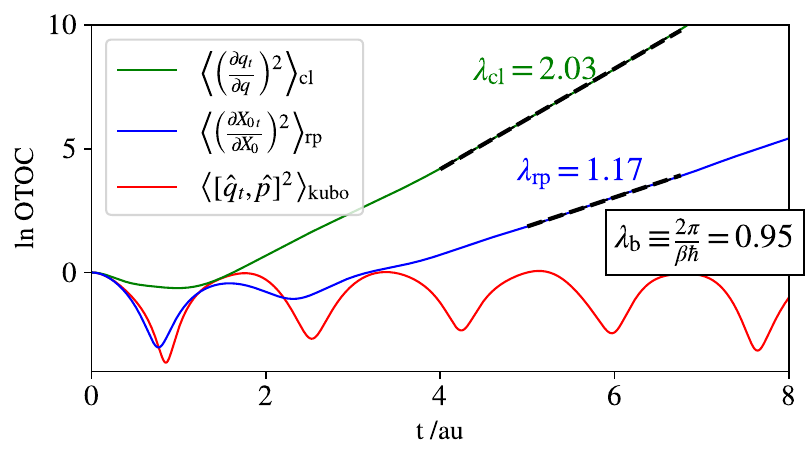}
}
\caption[Trivial counter-example of the RPMD bound violation]{A counter-example to show the early-time exponential growth of RPMD $\lambda_{\rm rp}$ violating the chaos bound $\lambda_{\rm b}$. Simulation and convergence parameters can be found in Appendix \ref{app:parametersforCounterExample}.}
\label{fig:counterexample}
\end{figure}

Here we define the two-dimensional classical Hamiltonian $H(\vc{p},\vc{q}) = T(\vc{p}) + V(\vc{q})$,  with
\begin{equation}
    T(p_x,p_y) = \frac{1}{2}(p_x^2+p_y^2)\text{,}
\end{equation}
and
\begin{equation}
\label{eqn:2dchaospotential}
    V(q_x,q_y) =  a(q_x^4 + q_y^4) + b(q_xq_y)^2 \text{.}
\end{equation}
Fig.~\ref{fig:2dqopot} shows that though there exists no stationary point in the potential, there are  large areas  of negative curvature, which may be adjusted by changing the ratio $a:b$.  
\\[5pt]
\indent
It is found that the values of $a=0.1$, $b=10$ are sufficient to show violation of the bound at $\beta = \frac{2\pi}{0.95}$, as shown in Fig.~\ref{fig:counterexample}. It should be noted that the early-time growth of the RPMD OTOC, $\lambda_{\rm rp}$, is still much smaller that that of the corresponding classical system, $\lambda_{\rm cl}$, indicating that although not sufficient to satisfy the bound, the effects of quantum thermal fluctuations contained in RPMD heavily damp the growth rate of OTOCs relative to the classical case.
\FloatBarrier
\begin{center}
\begin{tikzpicture}
\draw[thick] (0,0) -- (\linewidth,0);
\end{tikzpicture}
\end{center}
\FloatBarrier
\FloatBarrier
This result demonstrates that quantum detailed-balance is not a sufficient condition for satisfying the Maldacena bound. This is not especially surprising: although RPMD maintains correct quantum statistics, is a fictional dynamics, and the inclusion of springs introduces spurious artefacts, as commonly observed in RPMD infrared spectra \cite{rossiHowRemoveSpurious2014}. It is interesting to note that the quantum OTOC refuses to grow exponentially under these conditions, a common trait in the calculation of OTOCs in low-dimensional systems. This characteristic will be further discussed in Ch.\,\ref{5:}. 
\FloatBarrier
\subsection{Highly asymmetric barriers\label{Sec:AsymmetricBarriers}}
\begin{figure}[b!]
\centering
\scalebox{1}{\includegraphics[]{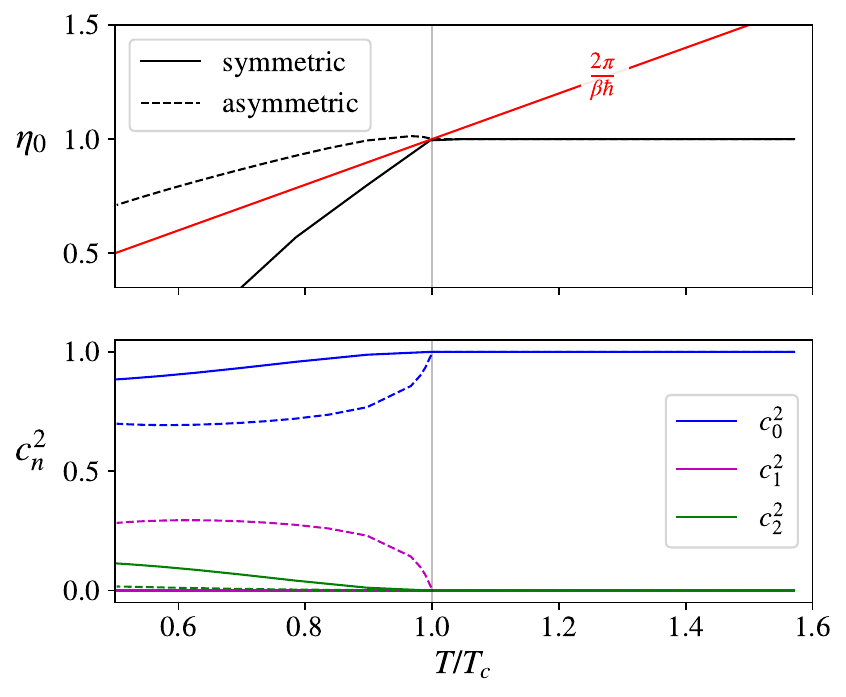}
}
\caption[Asymmetric instanton frequencies violate the bound]{The unstable mode frequency $\eta_0$ does not satisfy the chaos bound for asymmetric systems for $T\leq T_{\rm c}$ [top panel]. This is due to significant projection of the unstable mode into non-centroid modes [bottom panel]. Both models taken from Ref.\cite{richardsonRingpolymerMolecularDynamics2009}.}
\label{fig:eta0^2asymmVSsym}
\end{figure}

For asymmetric barriers at low temperatures there exists a significant projection of the unstable mode into the non-centroid modes of the ring polymer. We show an example of this in the lower panel of Fig.~\ref{fig:eta0^2asymmVSsym}, where we compare the projection of $\xi_0$ into $X_{|n|}$, quantified by $c_{n}$, for the asymmetric and symmetric Eckart models of Ref.\cite{richardsonRingpolymerMolecularDynamics2009}. For the symmetric case (solid lines), the projection of $\xi_0$ into the centroid (quantified by $c_0^2$) remains approximately $1$ for the entire range of temperatures, validating the approximation of Eq.\eqref{eqn:etaapproximation}. For the asymmetric model (dashed lines), this is not the case below $T_c$ as $c_{n>0}^2\not\approx0$.
\\[5pt]
\indent
As a consequence of the projection into the non-centroid modes, the instanton unstable mode frequency calculated for the asymetric model does not satisfy the chaos bound for $T<T_c$ (see dotted lines in the top panel of Fig.~\ref{fig:eta0^2asymmVSsym}), suggesting that the $t\to\infty$ limit of the RPMD OTOC will violate the bound.
\\[5pt]
\\[5pt]
\indent
It should be noted that we were unable to numerically verify the $t\!\to\!\infty$ RPMD OTOC growth directly in this section. This is due to the exponentially large number of trajectories required to sample the late-time growth making approach via direct calculation computationally infeasible. 
\section{Boltzmann averaging of unstable mode}
\begin{figure}[b!]
\centering
\scalebox{1}{\includegraphics[trim={1cm 0.45cm 0.35cm 0.45cm}]{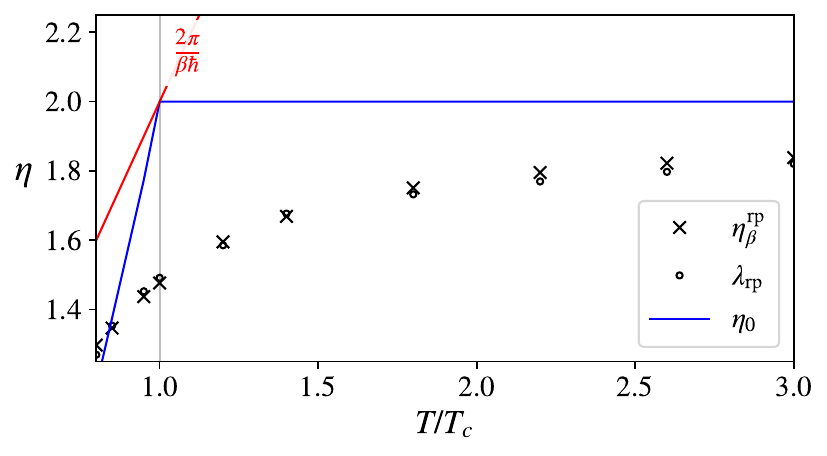}
}
\caption[Boltzmann averaging of the unstable mode in an asymmetric system]{Boltzmann averaging of the unstable mode for double well model of Ref.\cite{sadhasivamThermalQuenchingClassical2024}. Empty circles show early time RPMD OTOC growth for the same system from \cite{sadhasivamThermalQuenchingClassical2024}. Both results were reproduced with permission from this reference.}
\label{fig:sym_thermeta}
\end{figure}

It follows from Sec.\,\ref{Sec:AsymmetricBarriers} that due to projections of the unstable mode into the non-centroid modes of the instanton, the $t\to\infty$ growth of the RPMD OTOC will violate the Maldacena bound. 
\\[5pt]
\indent 
For early times, on the timescale on which quantum OTOCs grow exponentially, it is observed (compare $\eta_0$ with $\lambda_{\rm rp}$ in Fig.\,\ref{fig:sym_thermeta}) that the RPMD OTOC grows at a rate significantly lower than that of $\eta_0$. We suggest that is due to short-time thermal averaging of the unstable mode softening the early growth of the RPMD OTOC. 
\\[5pt]
\indent
To measure the thermal average of the unstable mode we define
\begin{equation}
    \eta_{\beta}^{\rm rp} := \sqrt{-\frac{1}{m}\left<\frac{\partial^2U_{N}(\vc{\tilde{q}})}{\partial \xi_0^2}\delta(\xi_{0})\right>_{\rm rp}} \text{,}
\end{equation}
where $\xi_0$ is the negative frequency normal mode of the ring polymer potential $U_N$ in the instanton configuration $\vc{\tilde q}$, and 
\begin{equation}
\label{eqn:rpAV}
    \left<\cdot\right>_{\rm rp} := \frac{1}{(2\pi\hbar)^{N}Z_N}\int \d\vc{p}\:\d\vc{q}\cdot e^{-\beta_N H_{N} (\vc{p},\vc{q})}
\end{equation}
is the ring polymer phase space average.
\\[5pt]
\indent 
In the following sections, we construct an expression for the unstable mode (following the work of Althorpe) then test whether this phenomenon is adequate to make RPMD OTOCs in asymmetric systems satisfy the bound on the timescales relevant to the quantum OTOC growth. 
\subsection{Geometry of the instanton and the unstable mode}
In the construction of the unstable mode, we consider the $N\to\infty$ limit of the ring polymer. In this limit the vector $\vc{q}$ becomes infinite dimensional, and thus can be equivalently represented as  $q(\tau)$, a periodic function of imaginary time $\tau$ with period $\beta\hbar$. If we select some point $0<\tau_0\leq\beta\hbar$ as the origin, may now consider the Fourier series corresponding to this periodic function,
\begin{equation}
\label{eqn:fourierexpansionofpath}
    q(\tau) = Q_0 + \sqrt{2} \sum_{n = 1}^{\infty}\left\{ Q_n (\tau_0)\sin\omega_{n} (\tau_0+\tau)  +  Q_{\bar n}(\tau_0) \cos\omega_n (\tau_0+\tau)\right\} \text{.}
\end{equation}
 Here, $\bar n = - n$ and $\omega_n = \frac{2\pi n}{\beta\hbar}$, the Matsubara frequencies of the system. It is important to stress that the Fourier modes $Q_n(\tau_0)$ do not have any explicit $\tau$ dependence, only dependence on the choice of origin $\tau_0$, with the centroid mode $Q_0$ being independent of this. 
 \\[5pt]
 \indent 
 The $N\to\infty$ limit of $U_N$ gives the Euclidean action of the loop in imaginary time,
\begin{equation}
\label{eqn:actionLOOP}
    S_E[\vc{q}] = \oint_{\beta\hbar} \left[\frac{m}{2}|\dot q(u)|^2 + V(q(u))\right] \d u \text{,}
\end{equation}
in which $\dot q \equiv \d q/\d \tau$. We now expand $S[\vc{q}]$ to second order about the instanton, $\tilde q(\tau)$ (or $\vc{\tilde Q}$ in the Fourier representation), in terms of the Fourier modes as
\begin{equation}
    S_E[\vc{q}] \simeq S_E[\vc{\tilde q}] + \frac{1}{2}\sum_{n,m\in\mathbb{Z}}(Q_n(\tau_0)-\tilde Q_n(\tau_0))\mt{\tilde K}(\tau_0)_{nm}(Q_m(\tau_0)-\tilde Q_m(\tau_0)) \text{,}
\end{equation}
where the instanton fluctuation matrix
\begin{equation}
    \mt{\tilde K}(\tau_0)_{nm} = m\omega_n^2\delta_{mn} + \frac{\partial}{\partial Q_n(\tau_0)}\frac{\partial}{\partial Q_m(\tau_0)}\oint_{\beta\hbar} V(\tilde q(u))\:\d u  \text{.}
\end{equation}
Here $\delta_{ij}$ is the Kronecker delta. The unstable mode corresponding to this choice of $\tau_0$, $\xi_0(\tau_0)$ is an eigenvector of $\mt{\tilde K}(\tau_0)$, with eigenvalue $-m\eta_0^2$.
\\[5pt]
\indent
For paths of stationary action, such as the instanton, Eq.\eqref{eqn:actionLOOP} generates\footnotemark{}\footnotetext{As the Principle of Least Action generates Hamiltonian dynamics in $\{p(t),q(t)\}$ for stationary $S:=\int [p^2/2m - V(q(t)]\d t$} the classical equations of motion of a particle moving on an upside-down potential, with position $q(\tau)$ at time $\tau$. The instanton can therefore be seen as a periodic orbit on the inverted potential with time period $\beta\hbar$, which is unstable in all but one degree of freedom. At the turning point in this trajectory the restoring force is therefore parallel to displacement, meaning the orbit must retrace itself. If the origin of imaginary time $\tau_0$ is fixed such that $\tilde q(\tau_0)$ is at the turning point of the orbit (i.e. setting $\tau_0=0$), 
\begin{equation}
\begin{split}
\label{eqn:symmetry_inst_asym}
    \tilde q(\tau) &= \tilde q(\beta \hbar - \tau)
    \\
    \implies \tilde Q_{n>0}(0) &= 0 {\text{,}}
\end{split}
\end{equation}
as the positive-$n$ Fourier modes, $Q_{n>0}(0)$ are all anti-symmetric with respect to imaginary time reversal. Given the choice of origin, $\tau_0=0$, the instanton must only occupy the space of the symmetric modes; $\{Q_0,Q_{n<0}(0)\}$. 
\\[5pt]
\indent 
The symmetry in Eq.\eqref{eqn:symmetry_inst_asym} also implies that $\mt{\tilde K}(0)$ is invariant to reflection $\tilde q(\tau) \leftarrow \tilde q(\beta \hbar - \tau)\:\forall\tau$, giving it a direct-product representation isomorphic to $C_S$. The unstable mode corresponding to this choice of imaginary time origin, $\xi_0(0)$ must contain $Q_0$ as these coordinates are identical for $T>T_{\rm c}$ (shown in Fig.~\ref{fig:eta0^2asymmVSsym} where $c_0^2=1$). Table~\ref{tab:irreps} shows the the representations of the Fourier modes in this point group, meaning that $\xi_0(0)$ must similarly only occupy the space of $\{Q_0,Q_{\bar n}(0)\}$. fluctuations about  $\tilde q$ in this direction therefore may be expressed as
\begin{equation}
\label{eqn:unstmode0}
    \xi_0(0) = c_0(Q_0-\tilde Q_0) + \sum_{n=1}^{\infty}c_n(Q_{\bar n}(0) - \tilde Q_{\bar n}(0)) \text{.}
\end{equation}
\indent
If $V(q)$ is also symmetric about its turning point, $\mt{\tilde K}(0)$ is now also invariant to $(q(\tau)\leftarrow -q(\tau))$, giving $C_{2v}$ symmetry. In this case, the even and odd modes are now also split into separate irreducible representations, as shown in Table~\ref{tab:irreps},
\begin{table}
\center{$\begin{array}{c|c|c|c|c|c|}
\cline{2-6}
\multicolumn{1}{c|}{}&Q_0 & Q_{\overline{n}\in{\rm even}}(0) & Q_{\overline{n}\in{\rm odd}}(0) & Q_{{n}\in{\rm even}}(0) & Q_{{n}\in{\rm odd}}(0) \\ \hhline{-|=|=|=|=|=|}
\multicolumn{1}{ |c||  }{C_{2v}}&A_1&A_1&B_1&A_2&B_2\\ \hline
\multicolumn{1}{ |c||  }{C_{S}}&A'&A'&A'&A''&A''\\ \hline
\end{array}$}
\caption[The symmetry of the normal modes]{The Fourier mode symmetry of $\tau_0$ = 0  instanton of $C_{2v}$ and $C_S$ symmetry}
\label{tab:irreps}
\end{table}
making the projection of $\xi_0(0)$ into $Q_{\bar n \in {\rm odd}}(0)$ also zero by symmetry (see for example $c_1^2$ in Fig.~\ref{fig:eta0^2asymmVSsym}). This lack of coupling is the reason that the assumptions of Eq.\eqref{eqn:etaapproximation} work so well for symmetric barriers.
\subsection{Permutational invariance\label{sec:perminvariance}}
To calculate of $\eta_{\beta}^{\rm rp}$ we average $\eta_0$ whilst constraining $\xi_0$, the unit-vector corresponding to movement through this negative curvature. In fulfilling this constraint, allowance for cyclic permutation of the loop must be made. Eq.\eqref{eqn:unstmode0}, gives the form of the unstable mode for the special case in which $\tau_0=0$. To generalise for all permutations, we use Eq.\eqref{eqn:fourierexpansionofpath} to translate the system from $\tau_0 = 0 \to \tau$.
\begin{equation}
\begin{split}
   \xi_0\left(\tau\right) = c_0(Q_0 - \tilde Q_0) &
   + \sum_{n=1}^{\infty} c_n\left\{(Q_{\bar n}(\tau) -  \tilde Q_{\bar n}(\tau))\cos \omega_n\tau\right.
   \\
   & \quad\quad\quad\quad\quad\quad\quad\quad - (Q_{n}(\tau) - \left.\tilde Q_{n}(\tau))\sin \omega_{n}\tau \right\} \text{.}
    \label{eqn:formofxi}
\end{split}
\end{equation}
We now define a set of permutationally-invariant coordinates $\{R_n\}$ such that
\begin{equation}
    \begin{split}
        Q_n(\tau) &= -R_n\sin \omega_n\tau\:\:\:\: \text{  and}\\
        Q_{\bar n}(\tau) &= R_n\cos \omega_n\tau\text{.}
    \end{split}
\end{equation}
In terms of these new coordinates,
\begin{equation}
\label{eqn:unstmode}
     \xi_0 = c_0(Q_0 - \tilde Q_0) 
   + \sum_{n=1}^{\infty}c_n(R_n - \tilde R_n) \text{,}
\end{equation}
in which $R_n^2 = Q_n(\tau)^2 + Q_{\bar n}(\tau)^2$. Note that unlike  the expression in Eq.\eqref{eqn:unstmode0},  Eq.\eqref{eqn:unstmode} is independent of $\tau_0$; it has been made permutationally invariant. 
\\
\indent 
For the special case of $c_{n>1} = 0$, level-sets of $\xi_0$ define cones in the space of $\{Q_0,Q_1,Q_{\bar 1}\}$; the approximate dividing surface of Ref.\cite{richardsonRingpolymerMolecularDynamics2009}. To calculate the $\eta_0$ at a certain configuration, we note that the transformation between $\{\vc{Q}\}$ to $\{\vc{\xi}\}$ was orthogonal and use the chain rule for derivatives (see App.~\ref{A:unstablemodederivs} for a more detailed discussion).
We now use the SHAKE algorithm \cite{ryckaertNumericalIntegrationCartesian1977} to thermally average the unstable mode whilst fulfilling the holonomic constraint defined by Eq.\eqref{eqn:unstmode}.
\subsection{Results}
\begin{figure}[]
\centering
\scalebox{1}{\includegraphics[trim={1cm 0.35cm 0.35cm 0.35cm}]{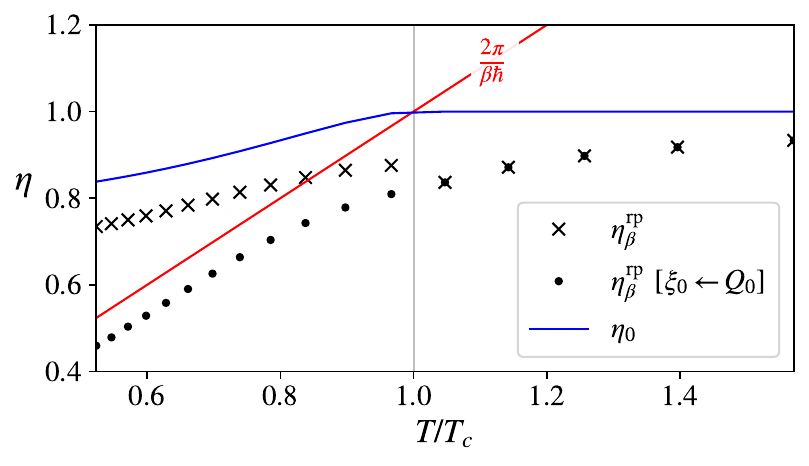}
}
\caption[Boltzmann averaging of the unstable mode in an asymmetric system]{Boltzmann averaging of the unstable mode in the asymmetric Eckart model of Ref.~\cite{richardsonRingpolymerMolecularDynamics2009}. Simulation details in App.~\ref{A:unstablemodederivs}.}
\label{fig:asym_thermeta}
\end{figure}

We cannot of course computationally implement the $N\to\infty$ limit of $\vc{q}$  as used in Sec.~\ref{sec:perminvariance}, but instead use $N$ as a convergence parameter. The frequencies of the Fourier modes are shifted for finite $N$ such that $\omega_n\leftarrow\omega_k$ where
\begin{equation}
    \omega_k = \frac{2}{\beta\hbar}\sin\left(\frac{k\pi}{N}\right)\quad\quad k = 0,1,\cdots N-1\text{.}
\end{equation}
As the coupling of $\xi_0$ into higher order fluctuations is negligible (shown in Fig.~\ref{fig:eta0^2asymmVSsym}) and $\omega_{k\ll N} \simeq \omega_{n=k}$ for large $N$,  we justify the usage of a finite, but large $N$.
\\[5pt]
\indent
To firstly show that $\eta_{\beta}^{\rm rp}$ is a good metric for the early-time growth of RPMD OTOCs, we compare with $\lambda_{\rm rp}$ for the symmetric double well of Ref.\cite{sadhasivamThermalQuenchingClassical2024}. Due to the minimal projection of $\xi_0$ into $Q_{n\neq0}$, we use a simple centroid constraint of $\xi_0\equiv X_0$. Fig.~\ref{fig:asym_thermeta} shows near-perfect agreement between the two, both above and below $T_c$, validating our approach. 
\\
\indent 
We now compute $\eta_\beta^{\rm rp}$ for the asymmetric model of Ref.\cite{richardsonRingpolymerMolecularDynamics2009} in which the unstable mode frequencies $\eta_0$ violated the Maldacena bound for $T<T_c$. Fig.~\ref{fig:asym_thermeta} shows that though thermal averaging does decrease the magnitude of $\eta_\beta^{\rm rp}$ relative to $\eta_0$, this is not sufficient for the satisfaction of the bound. We also plot the value of $\eta_\beta^{\rm rp}$ if the unstable mode were to be directed along the centroid, denoted as $\eta_\beta^{\rm rp} \:[\xi_0\leftarrow Q_0]$ in Fig.~\ref{fig:asym_thermeta}. These do satisfy the bound, showing again that the breakdown of the assumption of Eq.\eqref{eqn:etaapproximation} is responsible for violation of the Maldacena bound in RPMD.
\section{Conclusion}
 In this chapter we have shown that the infection of springs into ring polymer molecular dynamics can lead to violations of the Maldacena bound for RPMD OTOCs. The results of this chapter \textbf{do not contradict} the claim of Ref.\cite{sadhasivamInstantonsQuantumBound2023} that the Maldecena bound is enforced due to quantum Boltzmann statistics: thermal quenching of scrambling was consistently observed in the reduction of $\lambda_{\rm rp}$ versus $\lambda_{\rm cl}$ in all models considered. Rather, the results indicate that to properly test this theory, it must be done in a framework for which the underlying dynamics are accurate.
 \\[5pt]
 \indent
 The appropriate theory for this further work is Matsubara dynamics, but due to a severe sign-problem\cite{heleBoltzmannconservingClassicalDynamics2015}, there is no computationally feasible way to directly calculate a Matsubara OTOC. In the next chapter, we instead suggest an indirect method to calculate the late time growth of the Matsubara OTOC, avoiding this issue.

%% file: text/C2/C2+1.tex
\chapter{\label{4:}Matsubara OTOCs}
In the previous chapter we demonstrated that RPMD is an inadequate method to satisfy the Maldacena bound. In this chapter we illustrate the relationship between RPMD and Matsubara dynamics, then construct a method for the calculation of the late-time growth of a Matsubara OTOC. We focus only on systems at temperatures below $T_c$ as above this temperature all saddle-point systems trivially satisfy the bound.
\section{Relation of RPMD to Matsubara dynamics}
RPMD considers the dynamics of jagged ring polymers, for which the number of beads $N$ is a convergence parameter. Matsubara dynamics however describes that of smooth loops in imaginary time in the limit of $N\to\infty$, but with the number of Fourier modes (in this context referred to as Matsubara modes) having been kept at a finite $M$.  In this limit, the coordinates and momenta can, following the logic of Eq.\eqref{eqn:fourierexpansionofpath}, be written as a Fourier series in terms of the $M$ modes (which we assume to be odd) as
\begin{equation}
\label{eqn:matsq}
     q_{\rm mats}(\tau) = Q_0 + \sqrt{2} \sum_{n = 1}^{\mu}\left\{ Q_n (\tau_0)\sin\omega_{n} (\tau_0+\tau)  +  Q_{\bar n}(\tau_0) \cos\omega_n (\tau_0+\tau)\right\}\text{.}
\end{equation}
We remind the reader that though the non-centroid modes are independent of $\tau$, they do depend on the choice of imaginary time origin $\tau_0$. We have defined $\mu=(M-1)/2$ for notational convenience, and abbreviate the whole path as $\vc{Q}$, as knowledge of these $M$ modes specifies $q_{\rm mats}(\tau),\:\forall\tau$. The expression corresponding to $p_{\rm mats}(\tau)$ is analogous to Eq.\eqref{eqn:matsq}.
\\[5pt]
\indent
Following Willat\!\cite{willattMatsubaraDynamicsIts2017}\! we can analytically continue the Matsubara time-correlation function between $A(\vc{Q})$ and $B(\vc{Q})$ to give
\begin{equation}
\label{eqn:analyticallycontinuedMatsCqq}
    C_{AB}^{[M]} = \frac{1}{(2\pi\hbar)^MZ_M}\int\d\vc{Q}\,\d\vc{P}\: e^{-\beta R_{M}(\vc{Q},\vc{P})}A(\vc{Q})\:e^{\overline{\mathcal{L}}_Mt}B(\vc{Q})\text{,}
\end{equation}
where $R_{M}$ denotes the $N\to\infty$ limit of  the ring polymer Hamiltonian $H_N$, evaluated over the path defined by the Matsubara modes in Eq.\eqref{eqn:matsq}.
\begin{equation}
\label{eqn:Hntoinfty}
   R_M\equiv\lim_{N\to\infty} H_{N} =\oint_{\beta\hbar} \frac{p_{\rm mats}(\tau)^2}{2m}\d\tau + S_E[\vc{Q}]\text{.}
\end{equation}
Here, $U_N$ has become the Euclidean action $S_E$, as defined in Eq.\eqref{eqn:actionLOOP},  evaluated over the path $q_{\rm mats}(\tau)$.
\\[5pt]
\indent 
The only difference between Eq.\eqref{eqn:analyticallycontinuedMatsCqq} and the $N\to\infty$ limit of the corresponding RPMD correlation function is the Fourier smoothing of the paths integrated over (due to finite $M$), and the complex Liouvilllian
\begin{equation}
\label{eqn:matsubaraLiouvillian}
    \overline{\mathcal{L}}_M = \mathcal{L}_{\rm rp} + i\mathcal{L}_{\rm I} \text{.}
\end{equation}
The real part $\mathcal{L}_{\rm rp}$
\begin{equation}
    \mathcal{L}_{\rm rp} = \sum_{n=-\mu}^{\mu} \left\{\frac{P_n}{m}\frac{\partial}{\partial Q_n} - \frac{\partial S_E[\vc{Q}]}{\partial Q_n}\frac{\partial}{\partial P_n}\right\}\text{,}
\end{equation}
generates the classical dynamics of RPMD in the space of the $M$ Matsubara modes, whilst the imaginary part
\begin{equation}
    \mathcal{L}_{\rm I} = \sum_{n=-\mu}^{\mu}\omega_n\left(P_{\overline{n}}\frac{\partial}{\partial P_n} - Q_{\overline{n}}\frac{\partial}{\partial Q_n}\right)
\end{equation}
propagates trajectories into the complex plane, making the dynamics unstable and theory computationally infeasible in all but some applications \cite{pradaComparisonMatsubaraDynamics2023}.
\\[5pt]
\indent 
Removal of $\mathcal{L}_{\rm I}$ in analytically continued Matsubara therefore gives `smoothed RPMD', which suggests that some spurious artefacts observed in RPMD may be removed by inclusion of this extra part. In this chapter we consider these dynamics, when linearised about a fixed point, to obtain a late-time approximation to the growth of a Matsubara OTOC.
\section{Complex instantons\label{sec:complexinstantons}}
The complex Liouvillian cannot be written in terms of a Poisson bracket, hence there is no Hamiltonian associated with the complex dynamics and therefore the Matsubara phase space variables $\{\vc{Q},\vc{P}\}$ are not governed by Hamilton's equations of motion. Consequently, the intuition of `zero momentum + zero potential gradient $=$ fixed point' from classical mechanics is no longer applicable. Therefore, in this section we consider the action of $\overline{\mathcal{L}}_M$ on the stationary instanton configuration $\{\vc{Q},\vc{P}\}=\{\vc{\tilde Q},\vc{0}\}$ to see whether it remains a fixed point in analytically continued Matsubara dynamics. 
\\[5pt]
\indent 
We start by re-writing the Eq.\eqref{eqn:matsubaraLiouvillian} as
\begin{equation}
\label{eqn:CmplxL-II}
    \overline{\mathcal{L}}_M = \sum_{n=-\mu}^{\mu} \left\{\mathcal{L}_{Q_n} + \mathcal{L}_{P_n}\right\} + i\frac{\d}{\d\tau_0}\text{,}
\end{equation}
where we have used the chain rule and Eq.\eqref{eqn:matsq} to define the imaginry time translation operator
\begin{equation}
    \frac{\d}{\d\tau_0} = -\sum_{n=-\mu}^{\mu}\omega_n\left(P_{\overline{n}}\frac{\partial}{\partial P_n} + Q_{\overline{n}}\frac{\partial}{\partial Q_n}\right)\text{.}
\end{equation}
We have packaged up the other operators as
\begin{equation}
\label{eqn:LQn}
    \mathcal{L}_{Q_n} = \mathcolorbox{gray!15}{\frac{P_n}{m}}\,\frac{\partial}{\partial Q_n}\text{,}
\end{equation}
and
\begin{equation}
\label{eqn:LPn}
    \mathcal{L}_{P_n} = \mathcolorbox{gray!15}{\left(-\frac{\partial S_E[\vc{Q}]}{\partial Q_n}+2i\omega_nP_{\overline{n}}\right)}\,\frac{\partial}{\partial P_n}\text{.}
\end{equation}
This definition of Eq.\eqref{eqn:CmplxL-II} shows the Liouvillian as being made up of three types of operation. The first two, $\mathcal{L}_{Q_n}$ and $\mathcal{L}_{P_n}$, when applied on any function involve the multiplication by a prefactor (highlighted with \colorbox{gray!15}{grey boxes} for clarity) which when evaluated at the stationary instanton equal zero.  For the last operation however, this is not necessarily so. We may illustrate this by applying it on any non-centroid position mode, for example
\begin{equation}
\label{eqn:whizzingoffintoC}
    i\frac{\d}{\d\tau_0}Q_1 = -i\omega_1Q_{\bar{1}}\text{,}
\end{equation}
which is evidently not zero in the instanton configuration. We therefore conclude that, unlike in RPMD, the instanton in Matsubara dynamics is not stationary, as the position modes `whizz off' into the complex plane when evolved in time. 
\\[5pt]
\indent 
We now define imaginary-time translation invariant (ITTI) properties as some function of the phase space variables  $g(\vc{P},\vc{Q})$ for which
\begin{equation}
    \frac{\d}{\d\tau_0}\:g(\vc{P},\vc{Q}) = 0\text{.}
\end{equation}
An example of an ITTI coordinate already encountered in this report is $R_n$ as seen in Sec.\,\ref{sec:perminvariance}. In App.~\ref{app:ITTIproof}, we show that application of $\overline{\mathcal{L}}_M$ conserves imaginary time translation invariance of functions, meaning that all ITTI properties of the instanton are stationary in time under the dynamics of $\overline{\mathcal{L}}_M$. With this property in mind, we define the coordinate transformation from the Fourier modes to the `angular modes'
\begin{equation}
    \{\vc{P},\vc{Q}\}\to\{P,\vc{\Pi},\vc{L},Q,\vc{R},\vc{\Phi}\}\text{.}
\end{equation}
Here the centroid variables are unchanged ($\{P_0,Q_0\}\to\{P,Q\}$) and the fluctuation coordinates
\begin{equation}
    \begin{split}
        R_n &= \sqrt{Q_n^2+Q_{\overline{n}}^2}
        \\
        \Phi_n &= \sum_{k=1}^{\mu}T_{nk}\phi_k \text{,}\quad\quad\quad\quad\left[ \phi_n = \frac{1}{\omega_n}\arctan\left(\frac{Q_n}{Q_{\overline{n}}}\right)\right]
    \end{split}
\end{equation}
with $T_{nk}$ as the $\mu\times \mu$ discrete Fourier transform matrix as defined in Eq.\eqref{eqn:discretefourierMatrix}. The conjugate angular and radial momenta to $\{\vc{\Phi},\vc{R}\}$ are  $\{\vc{L},\vc{\Pi}\}$. 
\\[5pt]
\indent 
It can be shown that every single angular coordinate and conjugate momentum (except $\Phi_{\!0}$, the totally symmetric linear combination of $\phi_n$'s) is ITTI and therefore is stationary in the instanton configuration. As $\overline{\mathcal{L}}_M$ preserves ITTI symmetry, it maps all ITTI functions to those of the same symmetry. This means that the dynamics of coordinates of this symmetry are also all independent of $\Phi_{\!0}$.
\\[5pt]
\indent 
The dynamics of $\Phi_{\!0}$ can be understood by applying the Liouvillian
\begin{equation}
    \overline{\mathcal{L}}_M \Phi_{\!0} = -i\sqrt{\mu} +\frac{1}{\sqrt{\mu}}\sum_{n=-\mu}^{\mu}\frac{L_n}{mR_n^2}\text{,}
\end{equation}
where the conjugate momenta to $\phi_n$
\begin{equation}
    L_n = \frac{1}{\omega_n}\left(P_nQ_{\overline{n}}-P_{\overline{n}}Q_n\right)\text{.}
\end{equation}
One can show that these are also ITTI, hence all further powers in $\left(\overline{\mathcal{L}}_M \right)^n\Phi_{\!0}(\vc{Q})$ are also invariant, meaning that they are not a function of $\Phi_{\!0}$. Changing $\Phi_{\!0}$ therefore has \textbf{no effect} on the dynamics of the other coordinates. The movement of $\Phi_{\!0}\leftarrow\Phi_{\!0}+\Delta$ also has no effect even on its own dynamics, other than a translation by $\Delta$ such that $\Phi_{\!0\,t}\leftarrow\Phi_{\!0\,t}+\Delta ,\:\forall t$. 
\\[5pt]
\indent
We conclude that the instanton is a fixed point in analytically continued Matsubara dynamics in all but its redundant coordinate $\Phi_{\!0}$ when represented in angular coordinates. In the next section we use this result to approximate the late-time growth of Matsubara OTOCs.
\section{Matsubara OTOCs at `fixed' points}
In angular coordinates, we define the Matsubara stability matrix as
\begin{equation}
    \mt{M}_{\rm mats}(t) := \frac{\partial\:\{P_t,\vc{\Pi}_t,\vc{L}_t,Q_t,\vc{R}_t,\vc{\Phi}_t\}}{\partial\:\{P,\vc{\Pi},\vc{L},Q,\vc{R},\vc{\Phi}\}}\text{.}
\end{equation}
Using the complex Liouvillian, we construct the equations of motion of $\mt{M}_{\rm mats}$ as
\begin{equation}
    \frac{\d}{\d t}\mt{M}_{\rm mats}(t) = \mt{A}_{\rm mats}(t)\mt{M}_{\rm mats}(t) \text{.}
\end{equation}
As the instanton is stationary in all but the redundant coordinate, the propagation matrix in the stationary instanton configuration $\mt{\tilde{A}}_{\rm mats}$ is again constant, allowing us to calculate the growth of the instanton stability matrix from its eigenvalues. It is important to note that if one were to make the same construction under Fourier mode coordinates, the propagation matrix would no longer be stationary in the instanton geometry, making the following analysis impossible. We demonstrate this construction and discuss the insights gained in Appendix~\ref{app:Meom_Mats_Derivation}.
\\[5pt]
\indent
Due to the redundant nature of $\Phi_{\!0}$, as discussed in Sec.\,\ref{sec:complexinstantons}, we know that $\widehat{\underline{\Phi_{\!0}}}$ must be an eigenvector of $\mt{A}_{\rm mats}$ with eigenvalue $\lambda_0=0$. We can therefore calculate the survival probability, just as in Sec.~\ref{sec:multidimensionalClassicalScrambling}, but now without constraining $\Phi_{\!0}$, as 
\begin{equation}
\begin{split}
\label{eqn:Pmats}
    P_{\rm mats}(t) &=\int\d|\overline{\vc{z}}|\prod_{i=1}^{M-1}S(|\overline{z}_{i}|)S(|\overline{z}_{i\,t}|)\text{,}
\end{split}
\end{equation}
where $\{\overline{z}\}$ are the eigenvectors of $A$ with  $\widehat{\underline{\Phi_{\!0}}}$ removed. The fluctuations about the saddle obey equations of motion $\overline{z}_t = \overline{z}e^{\overline{\lambda}_it}$ where ${\overline{\lambda}_i}$ are the eigenvalues of $\mt{\tilde{A}}$ with $\lambda_0$ removed. This gives
\begin{equation}
\begin{split}
      P_{\rm mats}(t) &=  \int\d|\overline{\vc{z}}|\:\prod_{i=1}^{M-1}S(|\overline{z}_{i}|)S(|\overline{z}_{i}e^{\overline{\lambda}_it}|)
      \\
      &\sim \prod_{\Re{\lambda_i}>0}e^{-\lambda_it}\text{,}
\end{split}
\end{equation}
meaning that the late-time growth of $C_{\rm mats}$ is 
\begin{equation}
\begin{split}
\label{eqn:matsGrowth}
    C_{\rm mats}(t\to\infty) &\sim e^{2\max(\lambda)t}\prod_{\Re{\lambda_i}>0}e^{-\lambda_it}\text{,}
\end{split}
\end{equation}
which has the same form as the expression as for the multidimensional classical case. In the next section, we construct $\mt{\tilde{A}}_{\rm mats}$ for $M=3$ ($\mu=1$).
\section{3-mode Matsubara}
In angular coordinates, we define the $3$-Mode stability matrix as
\begin{align}
\label{eqn:MatsubaraM3}
    \mt{M_3} = \frac{\partial\{P_t,\Pi_t,L_t,Q_t,R_t,\Phi_t\}}{\partial\{P,\Pi,L,Q,R,\Phi\}}\text{,}
\end{align}
where all subscripts now indicate time dependence.
Applying $\overline{\mathcal{L}}_M$ onto Eq.\eqref{eqn:MatsubaraM3} and evaluating the result at the stationary instanton configuration $\{\vc{\tilde{Q}},\vc{0}\}$ gives the propagation matrix
\begin{align}
\label{eqn:MatsubaraA_3}
    \mt{\tilde A}_3 = 
    \left( \begin{array}{c c c|c c c}
    0&0&0&-{\frac{\partial^2S_E[\vc{Q}]}{\partial \tilde Q^2}}&-\frac{\partial^2S_E[\vc{Q}]}{\partial \tilde R\partial \tilde Q}&0\\
         0&0&-\dfrac{2i\omega_1}{\tilde R}&-\frac{\partial^2S_E[\vc{Q}]}{\partial \tilde Q\partial \tilde R}&-\frac{\partial^2S_E[\vc{Q}]}{\partial \tilde \tilde R^2}&0\\
         0&2i\omega_1 \tilde R&0&0&0&0\\
         \midrule
         \dfrac{1}{m}&0&0&&&\\
         0&\dfrac{1}{m}&0&&\mathlarger{\mathlarger{\mathlarger{\mathlarger{\vc{0}}}}}\quad\quad\quad\quad&\\
         0&0&\dfrac{1}{m\omega_1\tilde R^2}&&&\\
\end{array} \right)\text{.}
\end{align}
This matrix looks very similar to that of the multidimensional classical case of Eq.\eqref{eqn:classicalgeneralA}, save two key differences: first, the additional imaginary terms resulting from the imaginary part of the Liouvillian and second, the zeros in the Hessian block (upper-right $3\times3$ block in Eq.\eqref{eqn:MatsubaraA_3}). These zeroes are due to $\frac{\partial S_E}{\partial \Phi}=0$ as a consequence of the permutational invariance of the loop integral in Eq.\eqref{eqn:actionLOOP}.
\\[5pt]
\indent
As projection of the instanton configuration into modes of larger $|n|$ occurs at lower temperatures, similar to the unstable mode projection into non-centroid modes in Fig.\,\ref{fig:eta0^2asymmVSsym}, we conclude that temperatures as close to $T_c$ as possible are most appropriate for the $3$-mode approximation. 
\begin{figure}[]
\centering
\includegraphics[]{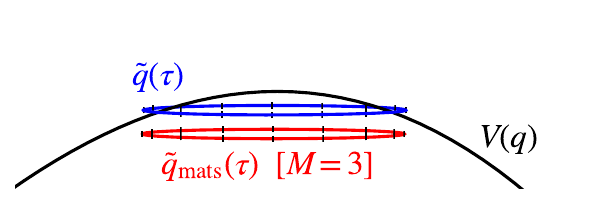}
\caption[Fitting of a $3$-mode Matsubara instanton to ]{Fitting of a 3-mode Matsubara loop $\tilde q_{\rm mats}(\tau)$ to the instanton loop $\tilde q(\tau)$.  $T/Tc = \frac{2\pi}{6.3} \approx 0.997$. $V(q)$ is the asymmetric Eckart model used in Ref.\cite{richardsonRingpolymerMolecularDynamics2009}. Black ticks are given to show imaginary intervals of $\beta\hbar/16$.}
\label{fig:3modefitting}
\end{figure}
\\[5pt]
\indent 
In Fig.\,\ref{fig:3modefitting} we compare the fit between the instanton at $T\approx0.977T_c$ for the asymmetric Eckart model of Ref.\cite{richardsonRingpolymerMolecularDynamics2009} and its $3$-mode Matsubara counterpart. We then calculate the derivatives of $S_E[\vc{Q}]$ for this configuration and diagonalise $\mt{\tilde A}_3$. For a fair comparison with RPMD, we setup the same 3-mode matrix, but without imaginary contributions and follow the same method.
\FloatBarrier
\begin{table}[H]
\center{$\begin{array}[]{c|c|c|c|c|c|c||c|}
\cline{2-8}
\multicolumn{1}{c|}{}&\lambda_1 & \lambda_2 & \lambda_3 & \lambda_4 & \lambda_5&\lambda_6& {\rm growth\: rate}\\ \hhline{-|=|=|=|=|=|=|=|}
\multicolumn{1}{ |c||  }{\rm Mats.}&1.9943&-1.9943&0.9946&-0.9946&0.0000&0.0000&0.9997\\ \hline
\multicolumn{1}{ |c||  }{\rm RPMD}&1.0000&-1.0000&0.1095i&-0.1095i&0.0000&0.0000&1.0000\\ \hline
\end{array}$}
\caption[3-mode max ]{The RPMD and Matsubara eigenvalues (to 5 significant figures) of the $3$-mode stability matrix propagator at the instanton for the asymmetric Eckart barrier of Ref.\cite{richardsonRingpolymerMolecularDynamics2009} at $T=0.997T_c$. Growth rates are then calculated using Eq.\eqref{eqn:matsGrowth}}
\label{tab:3moderesults}
\end{table}
The results of Table~\ref{tab:3moderesults} show large differences between the nature of the instanton in Matsubara and RPMD. For RPMD, as predicted by Eq.\eqref{eqn:relationoflambdatoeta}, we have: a pair of real eigenvalues $\lambda_1$ and $\lambda_2$, which correspond to motion along the unstable mode, a pair of imaginary ones $\lambda_3$ and $\lambda_4$ which correspond to stable motion orthogonal, and the final pair which we attribute to be due to the redundant coordinate. For 3-mode Matsubara though, as there are two pairs of real eigenvalues, the instanton now seems to behave as a second order saddle, showing that the inclusion of the imaginary terms in the Liouvillian makes major changes to the dynamics about this fixed point. Despite these discrepancies, the growth rates are approximately the same, the reason for which remains unclear.
\section{Conclusion}
Given that the 3-mode model is an extremely simplistic representation of an instanton, it is not appropriate at this point to draw any concrete conclusions about whether Matsubara OTOCs will adhere to the Maldacena bound. Instead it is intriguing to note the significant differences in dynamics between the two sets of results due to the complex Liouvillian, motivating further work on the model.

%% file: text/C3/C3.tex
\chapter{\label{5:}Scattering OTOCs -- `SOTOCs'} 
Throughout the analysis of the previous chapters it was found that quantum OTOCs very rarely grow exponentially in the case of low-dimensional systems. The fast onset of coherence due to interference is often blamed for this, as quantified by the `Ehrenfest time' $\tau_{\rm ehr}$. In this chapter we look at the growth of OTOCs in systems for which $\tau_{\rm ehr}$ is infinite, specifically scattering systems. No comparable studies have been identified in existing literature, hence the methodology employed will also be a topic of discussion.
\section{The model}
For our unbounded system, we consider a one-dimensional scattering system with Hamiltonian
\begin{equation}
    \hat H = \hat H_0 + V(\hat q)\text{,}
\end{equation}
where the unperturbed Hamiltonian 
\begin{equation}
    \hat H_0 = \frac{\hat p^2}{2m}\text{,}
\end{equation}
and 
\begin{equation}
\label{eqn:vScattering}
            V(\hat q) =
     \begin{cases}
      0 &\quad\left|q\right| > \frac{L}{2} \:\:\left(= \frac{\lambda}{\sqrt{8g}}\right)\\
        g\hat q^4-\frac{\lambda^2}{4}\hat q^2  + \frac{\lambda^4}{64g} &\quad\left|q\right|\leq \frac{L}{2}\quad{\rm .}\\
     \end{cases}
\end{equation}
A plot of this potential energy surface is shown in the right panel of Fig.~\ref{fig:SOTOCinfinitetemp} (red), with a superimposed plot of the symmetric double well from which it is derived (dotted blue). We use $\lambda =2$ and $g=0.08$ throughout the chapter. From now on we refer to $V(q)$ of Eq.\eqref{eqn:vScattering} as the `scattering potential' and the double well potential as the `bounded potential'.
\\[5pt]
\indent 
Since the potential vanishes for $|q|>L/2$, the eigenstates in these regions must be linear combinations of the free particle wavefunction. We therefore use the Numerov method\cite{numerovNoteNumericalIntegration1927} to numerically calculate the wavefunctions within the box, taking the free particle wavefunctions as boundary conditions. This matching to these asymptotic solutions is referred to as using `scattering boundary conditions' in literature. This generates  $\{\bra{x}\ket{\psi_p}\}$ such that:
\begin{equation}
\label{eqn:numerovbasis}
    \hat{H} \ket{\psi_p} = \frac{p^2}{2m}\ket{\psi_p}\text{.} \quad \quad \forall p \in \mathbb{R} 
\end{equation} 
As there exist no bound states in the scattering system, the scattering states of Eq.\eqref{eqn:numerovbasis} forms a complete, orthogonal set\cite{manolopoulosChemicalReactionDynamics2012}.
\section{\label{sec:boundstateissues}Problems with matrix element calculation}
Due to the infinite extent of the scattering states, it is impossible to calculate matrix elements by `brute force' integration as was possible over a finite DVR grid  for bounded systems (outlined in App.~\ref{app:qmOTOCS}).
The matrix elements are singular, as shown by this example:
\\[5pt]
\indent
Consider calculation of the matrix element $\bra{\psi_p}\hat p\ket{\psi_p}$. This integral can be separated in $x$ as
\begin{equation}\label{eqn:matelt_split}
\begin{split}
    \bra{\psi_p}\hat p\ket{\psi_p} = -i\hbar \left(\int\limits_{\:\:|x| \leq L/2}\!\!\!\!\!\!\d x\: \phi_p^*(x) \frac{\rm d}{{\rm d}x} \phi_p(x) 
   \:\: + \!\!\int\limits_{\:\:\:|x| > L/2}\!\!\!\!\!\!\d x\: \phi_p^*(x) \frac{\rm d}{{\rm d}x} \phi_p(x)
    \right)\text{,}
\end{split}
\end{equation}
where we have defined
\begin{equation}
    \phi_p(x) = \bra{x}\ket{\psi_p} \text{.}
\end{equation}
The first integral of the two is finite and is much like that of the integral over a finite DVR grid. The second integral however diverges, since
\begin{equation}
    \phi_p(x>L/2) \sim e^{ipx/\hbar}\text{.}
\end{equation}
This is not surprising as plane wave states are not normalisable. 
\section{Using wavepackets to calculate SOTOCs}
The form of $\hat H_0$ allows us to calculate OTOCs analytically in regions of space for which $\hat H = \hat H_0$. For example,
\begin{equation}
\label{eqn:contributionsfarfrombarrier}
\begin{split}
    [\hat q_t,\hat p] &= e^{it\hat H/\hbar}[\hat q,\hat p]e^{-it\hat H/\hbar}\quad\quad[\hat H =\hat H_0]
    \\
    &= i\hbar \text{,}
\end{split}
\end{equation}
therefore for all possible regularizations, the quantum OTOC
\begin{equation}
\begin{split}
    \left<[\hat q_t,\hat p]^2\right> &= \hbar^2\text{.}\quad\quad\quad\quad\quad\quad [\hat H =\hat H_0]
\end{split}
\end{equation}
This means that if we construct a basis localised in $q$, we can trace over the basis functions that contribute to the growth of the OTOC (in the vicinity of the barrier) and leave out those that do not.
\\[5pt]
\indent 
The position eigenstates at $q$ in the position representation are
\begin{equation}
\label{eqn:deltaposnfxn}
\bra{x}\ket{q} = \delta(x-q)\text{,}
\end{equation}
and in the momentum representation is
\begin{equation}
    \bra{p}\ket{q} = \frac{1}{\sqrt{2\pi\hbar}}e^{-\frac{ip}{\hbar}q}\text{,}
    \label{eqn:deltafxnMomentum}
\end{equation}
giving a uniform distribution in $p$ (as expected from the Heisenberg uncertainty principle). The infinite range of momenta means any real time evolution on the initial states will result in a spread of the wavefunction throughout all space. 
\\[5pt]
\indent 
Fortunately, we are interested in thermal OTOCs, for which the Boltzmann operator damps contributions of high momentum states in the basis. We can calculate this damping of Eq.\eqref{eqn:deltaposnfxn} analytically for $|x|>L/2$ where $\hat H =\hat H_0$,
\begin{equation}
\label{eqn:imaginarytimemomentumcutoff_xrep}
    \bra{x}e^{-\frac{\lambda\hat H_0}{2}}\ket{q} = \frac{1}{2\pi\hbar}\sqrt{\frac{4m\pi}{\lambda}}e^{\:-\frac{m}{\lambda}\left(\frac{x-q}{\hbar}\right)^2} \text{,}
\end{equation}
which can be transformed into the momentum representation as
\begin{equation}
\label{eqn:imaginarytimemomentumcutoff}
    \bra{p}e^{-\frac{\lambda\hat H_0}{2}}\ket{q} = \frac{1}{\sqrt{2\pi\hbar}}e^{\:-\frac{ip}{\hbar}q - \frac{\lambda p^2}{4m}} \text{,}
\end{equation}
showing that the Boltzmann operator gives an exponential cutoff in momentum, giving initial states far from the barrier a finite `range'. We show an example of this in Fig.~\ref{fig:findingX} where the evolved wavepackets are unable to reach the barrier in the time $t$. This makes possible the setup of an OTOC calculation in which we only include the states that interact with the barrier.
\\[5pt]
\indent 
We therefore calculate the thermal `scattering OTOC' (SOTOC) as
\begin{equation}
\begin{split}\label{eqn:SOTOS_general}
    S_\lambda(t) 
    &=\frac{1}{Z_s}\int_{q\in X} \d q\: \bra{q}e^{-\frac{\lambda\hat H}{2}} [\hat q_t,\hat p]^\dagger e^{-(\beta-\lambda)\hat H}[\hat q_t,\hat p] e^{-\frac{\lambda\hat H}{2}}\ket{q} \text{.}
\end{split}
\end{equation}
The interval $X$ defines a partial trace over the position basis such that all contributions outside of it do not involve wavepackets hitting the barrier in the time interval of interest $[0,t_{\rm max}]$. We similarly define the partition function as
\begin{equation}
    Z_s := \int_{q\in X}\d q\bra{q}e^{-\beta\hat H}\ket{q}\text{.}
\end{equation}
The SOTOC is calculated using wavepackets as
\begin{equation}
\begin{split}
\label{eqn:ptp0OTOC_LR}
    S_{\lambda}(t) &=\frac{1}{Z_s}\int_{q\in X} \d q\:\left[ \bra{L_{\lambda q}(t)}\ket{L_{\lambda q}(t)} + \bra{R_{\lambda q}(t)}\ket{R_{\lambda q}(t)} \right.
    \\
   & \quad\quad\quad\quad\quad\quad\quad\quad\quad\quad\quad\quad\quad\quad\quad -\left. 2{\rm Re}\left[\bra{R_{\lambda q}(t)}\ket{L_{\lambda q}(t)}\right] \:\right]  
\end{split}
\end{equation}
where the evolved wavepackets
\begin{equation}
\label{eqn:Lpsiops}
    \ket{L_{\lambda q}(t)} := e^{-\frac{\beta-\lambda}{2}\hat H} \hat p\: e^{\frac{it}{\hbar}\hat H} \: \hat q\: e^{-\frac{it}{\hbar}\hat H} e^{-\frac{\lambda}{2}\hat H}\ket{q} 
\end{equation}
and
\begin{equation}
\label{eqn:Rpsiops}
    \ket{R_{\lambda q}(t)} :=  e^{-\frac{\beta-\lambda}{2}\hat H}e^{\frac{it}{\hbar}\hat H} \: \hat q\: e^{-\frac{it}{\hbar}\hat H} \:\hat p\:e^{-\frac{\lambda}{2}\hat H}\ket{q} \text{.}
\end{equation}
Because $e^{-\frac{\lambda}{2}\hat H}\ket{q}$ is localised in both position and momentum, the evolved wavepackets can be propagated numerically, without the need for an infinite grid. In App.\:\ref{app:WPpropagation} we give details of this procedure.
\\[5pt]
\indent
From Eq.\eqref{eqn:SOTOS_general}, we calculate the standard and symmetrised SOTOCs with $S_\beta$ and  $S_{\beta/2}$ respectively. By noting that $S_\lambda(t) = S_{\beta-\lambda}(t)$, we obtain the Kubo-regularised SOTOC as
\begin{equation}
    \begin{split}\label{eqn:Skubo}
    S_{\rm kubo}(t) &
    = \frac{2}{\beta}\int_0^{\beta/2} \d\lambda\: S_{\beta-\lambda}(t)
    \text{.}
    \end{split}
\end{equation}
\section{\label{sec:discretizingbases}Discretizing bases}
Eq.\eqref{eqn:SOTOS_general} involves an infinite, continuous range of states that theoretically must to be integrated over. In this section, we present the necessary approximations required to make basis discrete, allowing integration over a finite grid making the method computationally feasible.
\\[5pt]
\indent 
We first discretize the eigenbasis $\{\ket{\psi_p}\}$ by adopting a discrete spectrum spaced by $\delta_p$ and introducing a maximum momentum cutoff $p_{\rm max}=N_p\delta_p$,  
\begin{equation}
\begin{split}
    \bra{x}\1\ket{x'} &= \int \d p \bra{x}\ket{\psi_p}\bra{\psi_p}\ket{x'} 
    \\
    &\simeq
    \sum_{n=-N_p}^{N_p} \phi_{n\smash{\delta_p}}(x)\,\phi_{n\smash{\delta_p}}^*(x')
    \delta_p
    \text{.}
\end{split}
\end{equation}
To discretise $\{\ket{q}\}$, the localised basis over which the SOTOC is integrated, we use 
\begin{equation}
\begin{split}
    \bra{x}\1\ket{x'} &= \int \d q \bra{x}\ket{q}\bra{q}\ket{x'}
    \\
    &\simeq 
    \sum_{n=-N_q}^{N_q}\chi_n(x)\,\chi_n^*(x') \delta_q
    \text{,}
\end{split}
\end{equation}
where we have, following Miller \etal\!\!\cite{colbertNovelDiscreteVariable1992a} defined the orthonormal sinc functions
\begin{equation}
    \chi_{n}(x) \equiv \bra{x}\ket{s_n} = \frac{\sin(\pi(x-n\delta_q)/\delta_q)}{\pi(x-n\delta_q)}\text{.}
 \end{equation}
We can further justify the usage of sinc functions by transforming them into the momentum representation,
\begin{equation}
    \chi_n(p) = \frac{1}{\sqrt{2\pi\hbar}}\:e^{-\frac{ip}{\hbar} n\delta_q}\:\Theta\left(\frac{p\,\delta_q}{2\pi\hbar}\right)\text{,}
\end{equation}
where the rectangular cutoff function
\begin{equation}
\Theta(x) = \left\{\begin{array}{ll} 0 & \quad\quad |x|>1/2
 \\1 & \quad\quad |x|\leq1/2
\end{array}   \right. 
\end{equation}
Comparison with Eq.\eqref{eqn:deltafxnMomentum} shows that the sinc function can be seen as a delta function with a momentum cutoff. This cutoff, along with $p_{\rm max}$ is suitable because states of maximal momenta are exponentially suppressed due to imaginary time propagation of the Boltzmann operator, as described in Eq.\eqref{eqn:imaginarytimemomentumcutoff}.
\section{\label{sec:SOTOCresults}Results}
\begin{figure}[b!]
\scalebox{1}{\includegraphics[trim={0.5cm 0.35cm 0.35cm 0.35cm}]{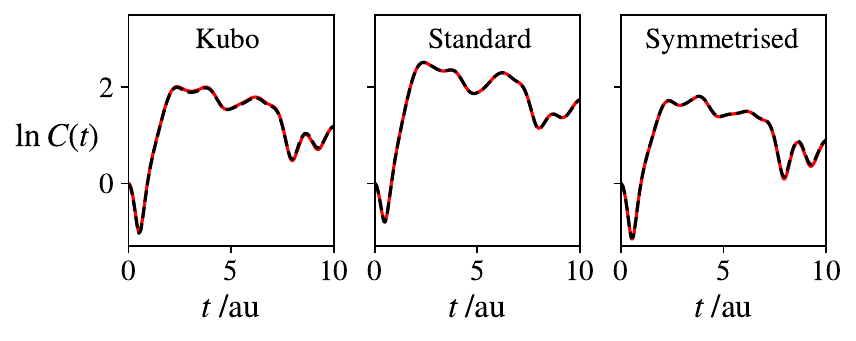}}
\caption[Comparison of wavepacket and matrix element method for a bounded system]{Comparison of the $[\hat q_t,\hat p]$ OTOC with matrix elements (red) and sinc wavepackets (dotted black line) for the bounded potential at $T=3T_c$. Simulation parameters in App.~\ref{app:WPpropagation}.}
\label{fig:wavepacketvalidation}
\end{figure}

\begin{figure}[t]
\scalebox{1}{\includegraphics[]{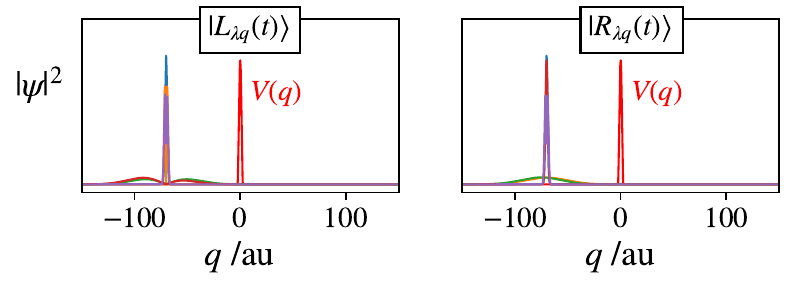}}
\caption[Calculation of the interval $X$]{Wavepacket propagation used to calculate the interval $X$. The left and right tiles display normalised probability distribution functions of the wavepackets for successive operator propagation as defined in Eq.\eqref{eqn:Lpsiops} and Eq.\eqref{eqn:Rpsiops}.}
\label{fig:findingX}
\end{figure}

As a test of the method, we used the wavepackets to calculate the OTOC for the bounded potential. For the example of $T=3T_c$ (and in all other temperatures checked), Fig.~\ref{fig:wavepacketvalidation} shows good agreement betweem the OTOC calculated using matrix elements, and using wavepackets, validating the usage of sinc functions as a complete basis.
\\[5pt]
\indent 
With this established, we calculate the SOTOC corresponding to the unbounded potential. To define the interval $X$ over which we put the sinc basis we specify the maximum time $t_{\rm max}=10$\,au. We then propagate wavepackets due to Eq.\eqref{eqn:Lpsiops} and Eq.\eqref{eqn:Rpsiops}, to find the minimum starting distance required for no overlap between the packet and the potential. In Fig.~\ref{fig:findingX}, we plot successive\footnotemark{} wavefunctions after each operator has been applied and find that for this choice of $t_{\rm max}$ and $T=3T_c$, a minimum range of $X=[-80,80]$ is required.  The OTOC corresponding to the propagation in Fig.~\ref{fig:findingX} gives the same value as for $\hat H \leftarrow \hat H_0$, further justifying this choice of $X$.
\footnotetext{ We do not label each wavefunction after each operator is applied in Fig.~\ref{fig:findingX} for brevity, as we are simply trying to demonstrate that the wavepacket does not hit the barrier at any point during propagation. }
\FloatBarrier
\begin{center}
\begin{tikzpicture}
\draw[thick] (0,0) -- (\linewidth,0);
\end{tikzpicture}
\end{center}
\FloatBarrier
\begin{figure}[H]
\scalebox{1}{\includegraphics[trim={0.5cm 0.35cm 0.35cm 0.35cm}]{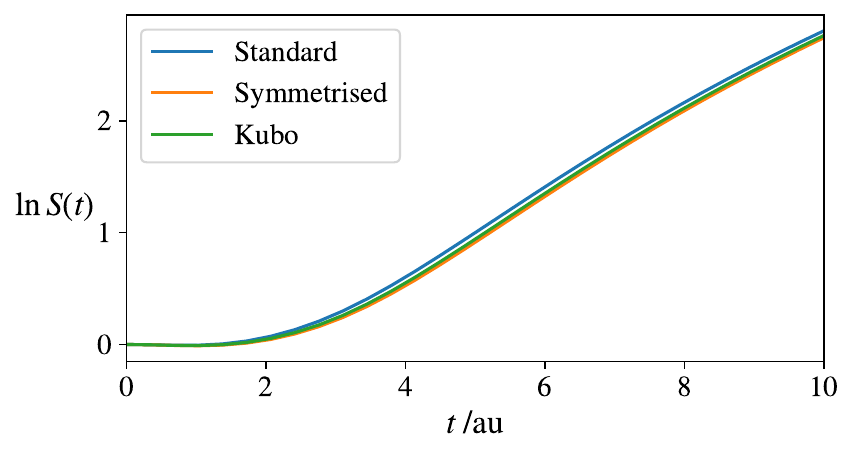}}
\caption[The ]{The $[\hat q_t,\hat p]$ SOTOC for the scattering system. $T=3T_c$. Further simulation details and  convergence parameters can be found in App.~\ref{app:WPpropagation}}
\label{fig:SOTOCresults}
\end{figure}
We show the SOTOC for the scattering potential in Fig.~\ref{fig:SOTOCresults} with $T=3T_c$. The results show exponential growth, but much less than that of the bounded OTOC of Fig.~\ref{fig:wavepacketvalidation}, which tapers off as $t$ approaches $10\,$au. Regularisation has essentially no effect on the SOTOC.
\begin{figure}[t]
\scalebox{1}{\includegraphics[trim={0.5cm 0.35cm 0.35cm 0.35cm}]{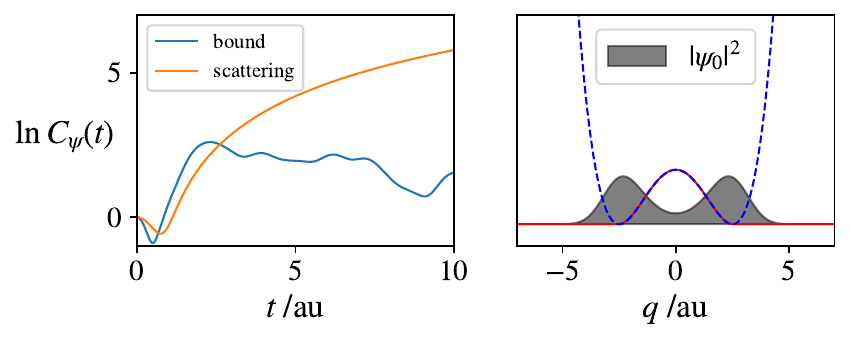}}
\caption[The ]{The infinite temperature $[\hat q_t,\hat p]$ OTOC with localised initial conditions. Further simulation details and  convergence parameters can be found in App.~\ref{app:WPpropagation}.}
\label{fig:SOTOCinfinitetemp}
\end{figure}
\\[5pt]
\indent 
To further verify this result, we calculate the infinite temperature ($\beta=0$) OTOC for both the bound and scattering systems. In this limit, the OTOC growth in both systems should be comparable, as the potential near the barrier and initial wavepackets are identical in both cases.  We can no longer use the sinc basis to approximate delta functions due to the lack of a thermal cutoff in momentum, as explained in Sec.~\ref{sec:discretizingbases}. We instead calculate the OTOC for a wavepacket defined by the minimum energy eigenstate of the bound system, which we denote as $C_{\psi}(t)$. The left panel of Fig.~\ref{fig:SOTOCinfinitetemp} shows an almost quantitative agreement in early time exponential growth rate between the scattering and bound systems, with the tapering off again observed in the scattering system for $t>4\,$au and the coherent oscillations for the bound system from $t>2.5\,$au.
\section{Conclusion}
In the OTOCs of scattering systems investigated, there is a noticeable decrease in the exponential growth as time progresses. As the Ehrenfest time is infinite in these systems, this flattening must result from some other interference effect in dynamics, which we suspect may be due to anharmonicities in the scattering potential. The high-temperature correspondence of Fig.~\ref{fig:SOTOCinfinitetemp} verifies the results of the SOTOC, and suggests that the reason for the meagre exponential growth of the thermal SOTOC in Fig.~\ref{fig:SOTOCresults} is due to the imaginary time propagation of the Boltzmann operator. However, it should be emphasised that the conclusions drawn here are somewhat unclear, and require further research to be verified.

%% file: text/C4/Conclusion.tex
\chapter{\label{6:}Conclusions and further work} 
In this report, we aimed to address two main questions. The first was to understand the influence of coherence on the flattening of quantum OTOCs. We compared the OTOCs of bounded and unbounded systems, outlining a methodology to calculate SOTOCs, thermal OTOCs of scattering systems specifically focusing on the contribution of wavepackets interacting with the potential barrier. We found that the exponential growth rate of the SOTOC was significantly lower than that of the analogous bounded system, which, following comparison with the results of the same systems at infinite temperature, we concluded was due to the action of the Boltzmann operator. For the OTOCs of unbounded systems we also observed a decrease in growth rate as time progressed, which we expect is consequence of the highly anharmonic scattering barrier, but require further research for verification. Though these results indicate that the Ehrenfest time is not solely responsible for the flattening of OTOCs, the precise mechanism still remains unclear, and will require further investigation to be understood.
\\[5pt]
\indent
The second question addressed the relationship between the Maldacena bound and the instanton, and whether RPMD is sufficient to satisfy the bound. The results of Chapter~\ref{3:} revealed that for systems lacking instantons RPMD produced OTOCs that, despite growing at a significantly lower rate than in the classical case, may still fail to satisfy the bound. In addition to this it was shown that some asymmetric systems for which instantons do form may also fail to satisfy the bound, even in the short-time limit when thermal quenching of the OTOC growth is significant. These results together provide overwhelming evidence that RPMD is not guaranteed to satisfy the Maldacena bound: though the bound may be of quantum statistical origin, conservation of the quantum Boltzmann distribution is not sufficient for it to be obeyed.
\\[5pt]
\indent 
These results suggested that a more sophisticated approach with accurate dynamics, Matsubara dynamics, would be necessary to understand how the bound is enforced. In Chapter~\ref{4:} we outlined an approach for the calculation of Matsubara OTOCs, by linearising the analytically continued Matsubara dynamics about the instanton. In considering the dynamics of the instanton itself, we uncovered that, unlike in RPMD, the instanton is not a fixed point, but instead is stationary in all coordinates but the collective angular coordinate $\Phi_0$. 
The minimal working example we tested for Matsubara OTOCs revealed dynamics that were markedly different to those of RPMD, showing a new perspective on the instanton that has not been seen before in literature. The theory presented in Chapter~\ref{4:} enables the developments of more complicated ($M>3$) Matsubara OTOC models, which we believe could shed light on the origin of the Maldacena bound in barrier scrambling systems and may also provide insights for the general case.
\\[5pt]
\indent
Just as the instanton governs the growth of OTOCs in barrier scrambling systems, it also controls the rate of quantum tunneling over barriers\cite{richardsonRingpolymerMolecularDynamics2009}, and is therefore the backbone of several reaction rate theories. This suggests that the further insight gained in this report regarding the nature of the instanton and dynamics of fluctuations about it could help inspire the construction of new rate theories, and aid further understanding of existing ones, providing a promising direction for further research.

%% file: text/CounterExParameters.tex
\chapter{\label{app:parametersforCounterExample}Simulation details for OTOC calculations}
In this section we give the parameters used for the calculation of OTOCs in Sec.~\ref{sec:classicalchaos}.
\section{RPMD and classical simulations\label{app:rpmdOTOCs}}
These simulations were split into two stages. The first was thermostatted RPMD to obtain an NVT ensemble. For all simulations, the thermostat of Ceriotti \etal in ref\cite{ceriottiEfficientStochasticThermostatting2010}. The thermostat parameters and duration of thermalisation were as follows
\begin{table}[H]
    \centering
    \begin{tabular}{|c|c|c|}
    \hline
       $\tau_0$  & $\lambda$ & $\tau_{\rm thrm}$\\
       \hline
        0.1 & 100&50~au\\
        \hline
    \end{tabular}
    \caption{Parameters for NVT sampling}
\end{table}
\noindent
The other parameters were as follows. For the RPMD results of Fig.~\ref{fig:counterexample}
\begin{table}[H]
    \centering
    \begin{tabular}{|c|c|c|c|}
    \hline
       $n_b$  & $dt$ & $n_{\rm traj}$&m\\
       \hline
         $64$&$0.01$~au &$1.5\times10^6$&$1$~au\\
         \hline
    \end{tabular}
    \caption{Miscellaneous parameters for RPMD OTOCs}
\end{table}
\noindent
for the Classical results
\begin{table}[H]
    \centering
    \begin{tabular}{|c|c|c|c|}
    \hline
       $n_b$  & $dt$ & $n_{\rm traj}$&m\\
       \hline
         $1$&$0.01$~au &$1.5\times10^6$&$1$~au\\
         \hline
    \end{tabular}
    \caption{Miscellaneous parameters for classical OTOCs}
    \label{tab:my_label}
\end{table}
\noindent 
The $4$\th order propagator of Manolopoulous \etal in Ref.\cite{brewerSemiclassicalDynamics151997} was initially used to propagate the Monodromy matrix, with a time step of $0.002{\rm au}$ from which the thermal OTOC was then constructed. We also verified out results using a finite difference approximation of the OTOC. In this method, we seed two identical trajectories for each inital condition in the NVT ensemble, perturbing one by a small parameter $\epsilon$. We then measure the stability matrix element using the ratio between the final difference and the initial difference of the phase space variables of interest. We found $\epsilon=1\times10^{-8}$ gave full convergence for all simulations.
\section{Quantum simulations\label{app:qmOTOCS}}
For the simulation of Quantum OTOCs, we used the discrete variable representation (DVR) basis of Colbert and Miller\cite{colbertNovelDiscreteVariable1992a} then numerically calculated matrix elements following the method of Hashimoto \etal in \cite{hashimotoOutoftimeorderCorrelatorsQuantum2017}.
For the DVR grid specifications, for the grid centered at $(0,0)$,
\begin{table}[H]
    \centering
    \begin{tabular}{|c|c|c|c|}
    \hline
        $L_x$ &  $L_x$ &$n_x$ &$n_y$\\
       \hline $6$~au &$6$~au &$100$&$100$\\
       \hline
    \end{tabular}
    \caption{Parameters for the DVR of the $2$D coupled quartic oscillators}
\end{table}
\noindent
A basis of $150$ eigenstates was generated, with $140$ of them used for insertion of the identity in matrix element calculation and $70$ traced over.

%% file: text/A-Shake.tex
\chapter{\label{A:unstablemodederivs}Thermal averaging of the unstable mode} 
In this appendix we outline the calculation of unstable mode potential energy curvature, and give simulation parameters for the SHAKE code.
\section{Calculating derivatives}
In the section we calculate the curvature of the potential energy in the direction of the unstable mode. As the transformation from $\{Q_0,R_1,...R_{\novtwo-1}\}$ to the unstable mode is orthogonal, the following holds:
\begin{equation}
    \frac{\rm d}{{\rm d}\xi_0} = c_0\frac{\partial}{\partial Q_0} +  \sum_{n \in m}c_n\frac{\partial}{\partial R_n} \text{,}
\end{equation}
where the set $m : \{1,...,\novtwo - 1\} $ and $R_n = \sqrt{Q_n^2 + Q_{\bar n}^2}$
Applying this once on $U$, the ring polymer potential gives,
\begin{equation}
    \frac{{\rm d}U}{{\rm d}\xi_0} = c_0\left(\frac{\partial U}{\partial Q_0}\right) + \sum_{n \in m}c_n\left(\frac{\partial U}{\partial R_n}\right) \text{,}
\end{equation}
and for a second time,
\begin{equation}
\begin{split}
    \frac{{\rm d}^2U}{{\rm d}\xi_0^2} &= c_0^2\left(\frac{\partial^2 U}{\partial Q_0^2}\right) + \sum_{n \in m }\!\!c_0c_n\!\left[\left(\frac{\partial^2 U}{\partial Q_0\partial R_n}\right) + \left(\frac{\partial^2 U}{\partial R_n\partial Q_0}\right)\right]       \\
    &\quad\quad\quad\quad\quad\quad\quad\quad\quad\quad\quad\quad\quad\quad\quad+ \sum_{n,n' \in m }\!\!\!c_n c_{n'}\left(\frac{\partial^2 U}{\partial R_n\partial R_{n'}}\right)\text{.}
\end{split}
\end{equation}
We then express the derivatives containing $R$ as those with $Q$. 
\begin{equation}
    \left(\frac{\partial^2 U}{\partial R_i\partial Q_j}\right) = \frac{Q_{i}}{R_i}\left(\frac{\partial^2 U}{\partial Q_i\partial Q_j}\right) + 
    \frac{Q_{\bar i}}{R_i}\left(\frac{\partial^2 U}{\partial Q_{\bar i}\partial Q_j}\right) \text{,}
\end{equation}
\begin{equation}
\begin{split}
 \left(\frac{\partial^2 U}{\partial Q_i\partial R_j}\right) &=
   \delta_{ij}\left[\frac{1}{R_j}\left(\frac{\partial U}{\partial Q_j}\right) -     \frac{Q_{j}^2}{R_j^3}\left(\frac{\partial U}{\partial Q_j}  \right)  - \frac{Q_{\bar j}Q_{j}}{R_j^3}\left(\frac{\partial U}{\partial Q_{\bar j}}  \right) \right]  
 \\
 & + \delta_{i\bar j}\left[\frac{1}{R_j}\left(\frac{\partial U}{\partial Q_{\bar j}}\right) -     \frac{Q_{\bar j}^2}{R_j^3}\left(\frac{\partial U}{\partial Q_{\bar j}}  \right)  - \frac{Q_{\bar j}Q_{j}}{R_j^3}\left(\frac{\partial U}{\partial Q_{j}}  \right) \right]
 \\
 & +  \frac{Q_j}{R_j}\left(\frac{\partial^2 U}{\partial Q_i\partial Q_j}\right) + 
    \frac{Q_{\bar j}}{R_j}\left(\frac{\partial^2 U}{\partial Q_i\partial Q_{\bar j}}\right) 
\end{split}
\end{equation}
and
\begin{equation}
\begin{split}
        \left(\frac{\partial^2 U}{\partial R_i\partial R_j}\right) = \frac{Q_{i}}{R_i}\left(\frac{\partial^2 U}{\partial Q_i\partial R_j}\right) + 
    \frac{Q_{\bar i}}{R_i}\left(\frac{\partial^2 U}{\partial Q_{\bar i}\partial R_j}\right) \text{,}
\end{split}
\end{equation}
from which the unstable mode derivatives may be extracted.
\section{Simulation parameters}
For thermal averaging of the unstable mode we use the same NVT/NVE method with the same type of thermostat, generating a constrained NVT ensemble, using SHAKE. To calculate the unstable mode coefficients, we cyclically permute the instanton such that $\tau_0=0$, then diagonalize the Hessian and take the eigenvector corresponding to the negative eigenvalue. For computational ease we introduced a cutoff on these coefficients at $0.001$ such that any below this value were treated as zero.
\\
\indent 
For constrained thermalization asymmetric double well, we use
\begin{table}[H]
    \centering
    \begin{tabular}{|c|c|c|c|}
    \hline
       $\tau_0$  & $\lambda$ & $\tau_{\rm thrm}$&dt\\
       \hline
        1 & 1&80~au&0.002~au\\
        \hline
    \end{tabular}
    \caption{Parameters for NVT sampling for the results of Fig.~\ref{fig:asym_thermeta}.}
\end{table}
\noindent
The other parameters were as follows
\begin{table}[H]
    \centering
    \begin{tabular}{|c|c|c|}
    \hline
       $n_b$  &  $n_{\rm traj}$&m\\
       \hline
         $256$&$5000$&$1$~au\\
         \hline
    \end{tabular}
    \caption{Miscellaneous parameters for SHAKE}
\end{table}

%% file: text/MeomMatsDerivation.tex
\chapter{Matsubara stability matrix propagation in Fourier modes\label{app:Meom_Mats_Derivation}}
In this appendix we consider the Matsubara stability matrix propagator for Fourier modes, to gain insight into the effects of imaginary part in the complex liouvillian.
From Eq.\eqref{eqn:matsubaraLiouvillian} we have the complex equations of motion of $\vc{P}$ and $\vc{Q}$ as
\begin{equation}
\begin{split}
    \frac{\d Q_n}{\d t} &= \frac{P_n}{m} - i\omega_nQ_{\overline{n}}\quad\quad\text{and}\\
    \frac{\d P_n}{\d t} &= -\frac{\partial U_N(\vc{Q})}{\partial Q_n} + i\omega_nP_{\overline{n}} \:\text{.}
\end{split}
\end{equation}
These we can write in a more compact form as
\begin{equation}
\begin{split}
    \frac{\d \vc{Q}}{\d t} &= \frac{\vc{P}}{m} - i\mt{U}\vc{Q}
    \quad\quad\text{and}\\
    \frac{\d \vc{P}}{\d t} &= -\frac{\partial U_N(\vc{Q})}{\partial \vc{Q}} + i\mt{U}\vc{P} \:\text{,}
\end{split}
\end{equation}
for which we have defined the adjacency matrix 
\begin{equation}
    \mt{U} = \begin{pmatrix}
        0 & \cdots&    \cdots &0   &\omega_{\overline{N}}\\
    \vdots&   0    & \iddots    &\iddots   & 0\\
    \vdots &\iddots &  \omega_0 & \iddots& \vdots\\
        0  &\iddots&\iddots &    0   &\vdots\\
  \omega_{N}&   0  &    \cdots&  \cdots &0
    \end{pmatrix}
\end{equation}
and packaged the phase-space variables as
\begin{equation}
    \vc{Q} = \begin{pmatrix}
        Q_{\overline{N}}\\
        \vdots\\
        Q_0\\
        \vdots\\
        Q_N
    \end{pmatrix}
    \quad\quad   \quad\quad   \quad\quad
        \vc{P} = \begin{pmatrix}
        P_{\overline{N}}\\
        \vdots\\
        P_0\\
        \vdots\\
        P_N
    \end{pmatrix}
\end{equation}
We now apply the chain rule on the derivative on the stability matrix elements $\mt{M_{XY}}\:\: [X,Y\in\{P,Q\}]$
\begin{equation}
    \frac{\d}{\d t}\frac{\partial \vc{X}_t}{\partial\vc{Y}} = \frac{\partial}{\partial \vc{Y}}\frac{\d \vc{X}_t}{\d t} \text{,}
\end{equation}
to yield
\begin{equation}
\label{eqn:MQQ}
    \frac{\d \mt{M_{QQ}}}{\d t} =\frac{1}{m}\mt{M_{PQ}} - i\mt{U}\mt{M_{QQ}}
\end{equation}
\begin{equation}
\label{eqn:MQP}
    \frac{\d \mt{M_{QP}}}{\d t} =\frac{1}{m}\mt{M_{PP}} - i\mt{U}\mt{M_{QP}}
\end{equation}
\begin{equation}
\label{eqn:MPP}
    \frac{\d \mt{M_{PP}}}{\d t} =-\frac{\partial^2 U_N(\vc{Q}_t)}{\partial \vc{Q}_t^2}\mt{M_{QP}} + i\mt{U}\mt{M_{PP}}
\end{equation}
\begin{equation}
\label{eqn:MPQ}
    \frac{\d \mt{M_{PQ}}}{\d t} =-\frac{\partial^2 U_N(\vc{Q}_t)}{\partial \vc{Q}_t^2}\mt{M_{QQ}} + i\mt{U}\mt{M_{PQ}}\text{,}
\end{equation}
which is then re-displayed in matrix form as
\begin{equation}
\label{eqn:Meom_matsubara}
    \frac{\d}{\d t} \underbrace{\begin{pmatrix}
        \mt{M_{QQ}} & \mt{M_{QP}} \\
        \mt{M_{PQ}} & \mt{M_{PP}}
    \end{pmatrix}}_{\mathlarger{\mt{M}_{\rm mats}^{\rm F}(t)}} = 
    \underbrace{\begin{pmatrix}
        -i\mt{U}  &  m^{-1}\mt{1}  \\
        -\mt{K}(\vc{q}_t,\tau_0) & i\mt{U}
    \end{pmatrix}}_{\mathlarger{\mt{A}_{\rm mats}^{\rm F}(t)}}
      \begin{pmatrix}
        \mt{M_{QQ}} & \mt{M_{QP}} \\
        \mt{M_{PQ}} & \mt{M_{PP}}
    \end{pmatrix}
   \text{.}
\end{equation}
Here, the Fourier Matsubara stability matrix stability matrices
\begin{equation}
    \mt{M}_{\rm mats}^{\rm F}(t) := \frac{\partial \{\vc{Q}_t,\vc{P}_t\}}{\partial \{\vc{Q},\vc{P}\}} \text{,}
\end{equation}
and the fluctuation matrix
\begin{equation}
    \mt{K}(\vc{q},\tau_0)_{nm} = m\omega_n^2\delta_{mn} + \underbrace{\frac{\partial}{\partial Q_n(\tau_0)}\frac{\partial}{\partial Q_m(\tau_0)}\oint_{\beta\hbar} V(q(u))\:\d u}_{\mathlarger{\mt{K}_{0}(\vc{q},\tau_0)_{nm}}}  \text{.}
\end{equation}
We define the matrix
\begin{equation}
    \mt{U}_{nm} = \omega_n\delta_{\overline{n}m}\text{,}
\end{equation}
and use $\mt{1}$ to denote an identity matrix the size of the stability matrix. We now manipulate Eq.\eqref{eqn:Meom_matsubara} to give
\begin{equation}
\label{eqn:Mqq_mats_evolution}
    \frac{\d^2}{\d t^2}\mt{M_{QQ}} = -\frac{1}{m}\mt{K}_{0}(\vc{q}_t,\tau_0)\mt{M_{QQ}}\text{,}
\end{equation}
noting that $\mt{M_{QQ}}_{00} \equiv \frac{\partial X_{\!0\:t}}{\partial X_{\!0}}$. Eq.\eqref{eqn:Mqq_mats_evolution} therefore shows an example of the growth of the Matsubara OTOC not having any springs. Another way to look at this is by using the fact that
\begin{equation}
    \left(\lambda_{k}\mt{1} + i\mt{U}\right)\frac{1}{m}\mt{1} = \frac{1}{m}\mt{1}\left(\lambda_{k}\mt{1} + i\mt{U}\right) \quad\quad\quad\forall k
\end{equation}
meaning that the eigenvalues of $\mt{A}_{\rm mats}^{\rm F}(t)$ which we denote as $\lambda_k, k\in\{\pm1,\pm2\cdots\pm M\}$ may be found using 
\begin{equation}
\begin{split}
\det{\left(\lambda_{k}\mt{1} + i\mt{U}\right)\left(\lambda_{k}\mt{1} - i\mt{U}\right)+\frac{1}{m}\mt{K}(\vc{q}_t,\tau_0)}=0\text{.}
\end{split}
\end{equation}
Now we note that $\mt{U}^2$ gives a diagonal matrix of $-\omega_n^2$, which remove the spring contributions of the RPMD Hessian,
\begin{equation}
\begin{split}
\therefore \quad \det{\lambda_{k}^2\mt{1}+\frac{1}{m}\mt{K}_0(\vc{q}_t,\tau_0)}=0 
\end{split}
\end{equation}
There are no spring contributions in the growth of Matsubara OTOCs. 
\\
\indent 
Above $T_c$, the saddle configuration $\vc{q}=0$ can be shown to be stationary for the analytically-continued propagator. Even though this is true, we may \textbf{not} expand about the Fourier modes when calculating the Matsubara stability matrix propagator as small fluctuations in the non-centroid coordinates do not preserve $\mt{\tilde A}_{\rm mats}^{\rm F}$ as they entail a descent in symmetry of the path. I think the only way to get the above $T_c$ expansion would be to setup the problem in angular modes, the take $R_n\to0$ and deal with the $Q_n/R_n$'s in the limit. 

%% file: text/ITTIstationaryProof.tex
\chapter{Proof of stationary ITTI coordinates\label{app:ITTIproof}}
We firstly, following Althorpe, show that derivative products of form
\begin{equation}
    d_1[fg] = \sum_{n=-\mu}^{\mu}\frac{\partial f}{\partial Q_n}\frac{\partial g}{\partial Q_n}
\end{equation}
or
\begin{equation}
    d_2[fg] = \sum_{n=-\mu}^{\mu}\frac{\partial f}{\partial Q_n}\frac{\partial g}{\partial P_n}
\end{equation}
is permutationally invariant (PI) provided $f$ and $g$ are PI. Explicitly for $d_1[fg]$ we have
\begin{equation}
    \frac{\d d_1[fg]}{\d\tau_0} = \underbrace{-\sum_{m=-\mu}^\mu\omega_m\left[Q_{\overline{m}}\frac{\partial}{\partial Q_m}+P_{\overline{m}}\frac{\partial}{\partial P_m}\right]}_{\mathlarger{\mathlarger{\frac{\d}{\d\tau_0}}}}\sum_{n=-\mu}^\mu\frac{\partial f}{\partial Q_n}\frac{\partial g}{\partial Q_n}\text{.}
\end{equation}
Now all of the derivatives in $d_1[fg]$ may be pulled out behind the derivatives of $\partial/\partial \tau_0$ , except those of $\partial/\partial Q_m$ for $m=\overline{n}$, due to the multiplication of $Q_m$. An example of this is
\begin{equation}
\label{app:extraterms}
    \omega_{\overline{n}}Q_n\frac{\partial }{\partial Q_{\overline{n}}}\left[\frac{\partial f}{\partial Q_{n}}\right] = \frac{\partial}{\partial Q_n}\left[\omega_{\overline{n}}Q_n\frac{\partial f}{\partial Q_{\overline{n}}}\right] \underbrace{- \omega_{\overline{n}}\frac{\partial f}{\partial Q_{\overline{n}}}}_{{\rm extra\:term}}
\end{equation}
with a similar result for $g$. As $f$ and $g$ are PI, their $\tau_0$ derivatives are zero, hence we have only the extra terms from Eq.\eqref{app:extraterms} left over such that,
\begin{equation}
    \begin{split}
        \frac{\d d_1[fg]}{\d\tau_0} &= -\sum_{n=-\mu}^\mu \omega_n \left[\frac{\partial f}{\partial Q_{\overline{n}}}\frac{\partial g}{\partial Q_{n}}+\frac{\partial f}{\partial Q_{n}}\frac{\partial g}{\partial Q_{\overline{n}}}\right]
        \\ &= -\sum_{n=-\mu}^\mu \omega_n \left[\frac{\partial f}{\partial Q_{\overline{n}}}\frac{\partial g}{\partial Q_{n}}-\frac{\partial f}{\partial Q_{\overline{n}}}\frac{\partial g}{\partial Q_{{n}}}\right] = 0\text{,}
    \end{split}
\end{equation}
where in the second line we used $\omega_{\overline{n}}=-\omega_n$ and noted that the sum goes over all pairs of $n$ and $\overline{n}$. The analogous proof holds for $d_2[fg]$.
\\
\indent 
The complex Liouvilllian
\begin{equation}
\label{app:matsubaraLiouvillian}
    \overline{\mathcal{L}}_M = \mathcal{L}_{\rm rp} + i\mathcal{L}_{\rm I} \text{,}
\end{equation}
with
\begin{equation}
    \mathcal{L}_{\rm rp} = \sum_{n=-\mu}^{\mu} \left\{\frac{P_n}{m}\frac{\partial}{\partial Q_n} - \frac{\partial S_E(\vc{Q})}{\partial Q_n}\frac{\partial}{\partial P_n}\right\}\text{,}
\end{equation}
and
\begin{equation}
    \mathcal{L}_{\rm I} = \sum_{n=-\mu}^{\mu}\omega_n\left(P_{\overline{n}}\frac{\partial}{\partial P_n} - Q_{\overline{n}}\frac{\partial}{\partial Q_n}\right)\text{.}
\end{equation}
We can therefore write the Liouvillian acting on some PI function $g$ as
\begin{equation}
\overline{\mathcal{L}}_Mg = d_2[gf_1] + d_2[f_2g] + i\left(d_2[f_3g] + d_2[gf_4]\right)
\end{equation}
where we have defined
\begin{equation}
\begin{split}
    f_1 &:= \sum_{n=-\mu}^\mu \frac{P_n^2}{2m}
    \\
    f_2&:= -S_E(\vc{Q})
    \\
    f_3 &:= \sum_{n=-\mu}^\mu \omega_nP_{\overline{n}}Q_n
    \\
        f_4 &:= - f_3 
\end{split}
\end{equation}
All of these functions can be shown to the PI, making $\overline{\mathcal{L}}_Mg$ also PI. Therefore the application of the liouvillian conserves permutational invariance of functions.
\\
\indent
To calculate the time propagation of some function os the variables at time $t$, we use
\begin{equation}
    f((\vc{P}_t,\vc{Q}_t);(\vc{P},\vc{Q})) = e^{\overline{\mathcal{L}}_Mt}f(\vc{P},\vc{Q})
\end{equation}
we then Taylor expand in $t$ to give
\begin{equation}
    f((\vc{P}_t,\vc{Q}_t);(\vc{P},\vc{Q})) = f(\vc{P},\vc{Q}) + {\sum_{n=1}^\infty \frac{1}{n!}\left({\overline{\mathcal{L}}_M}t\right)^nf(\vc{P},\vc{Q})}\text{,}
\end{equation}
where the second term is zero for PI $f$ at the stationary instanton configuration as discussed in Sec.~\ref{sec:complexinstantons}.

%% file: text/WavepacketCalculation.tex
\chapter{Wavepacket propagation \label{app:WPpropagation}}
Firstly, we outline the method used to numerically calculate the evolved wavepackets of Eq.\eqref{eqn:Lpsiops} and Eq.\eqref{eqn:Rpsiops}. Taking $\ket{L_{\lambda q}(t)}$ as an example, we integrate
\begin{equation} \label{eqn:LqIntegral}
\begin{split}
    \bra{\psi_p}\ket{L_q} =& -\hbar^2 e^{-\frac{\beta-\lambda}{2}E_p}\int \d x'' {\phi^*_p}(x'') \frac{\rm d}{{\rm d}x''}\int \d p'' \phi_{p''}(x'') \:e^{\frac{it}{\hbar}E_{p''}}
    \int \d x'  {\phi^*_{p''}}(x')
    \\
   &\quad\quad \times  \frac{\rm d}{{\rm d}x'} \int \d p' \phi_{p'}(x')\: e^{\frac{it}{\hbar}E_{p'}} \int \d x \:{\phi^*_{p'}}(x) \bra{x}\ket{q_\lambda} {\rm ,}
\end{split}
\end{equation}
starting from the innermost integral. Here $\bra{x}\ket{q_\lambda}$ is the thermally softened sinc function, like in Eq.\eqref{eqn:imaginarytimemomentumcutoff}. We numerically calculate all integrals using trapezium rule, converging with respect to grid length and spacing.
\section{Wavepacket validation parameters}
For the `matrix elements' calculation, the same method as in App.~\ref{app:qmOTOCS} was used, with DVR grid 
parameters
\begin{table}[H]
    \centering
    \begin{tabular}{|c|c|}
    \hline
        $L_x$  &$n_x$ \\
       \hline $12$~au &$2500$\\
       \hline
    \end{tabular}
    \caption{Parameters for the DVR of the $1$D double well}
\end{table}
\noindent 
A basis of $150$ eigenstates was generated, with $140$ of them used for insertion of the identity in matrix element calculation and $70$ traced over. The mass $m=0.5~{\rm au}$ for all simulations in the chapter.
\\
\indent 
The same eigenbasis was used for the wavepacket propagation, but now as a trace over $n_{\rm pkts}=50$ evenly spaced sinc functions over the length $L_{\rm basis}=15~{\rm au}$. For the Kubo regularised OTOC, the integral of Eq.\ref{eqn:Skubo} was calculated using trapezium rule for $N_{\rm kubo}=50$ evenly spaced points in $\lambda$.
\section{SOTOC parameters}
For the results of the scattering system, the basis was discretised as follows
\begin{table}[H]
    \centering
    \begin{tabular}{|c|c|c|c|}
       \hline $p_{\rm max}$  & $\delta_{p}$ & $L_{x}$&$n_x$\\
        \hline $28.619~{\rm au}$& $1.432\times10^-5~{\rm au}$&$500~{\rm au}$&$5000$\\ \hline
    \end{tabular}
    \caption{Parameters of the discretized basis used in scattering calculations}
\end{table}
\noindent
For the SOTOC calculation, the sinc function parameters were $L_{\rm basis}=160~{\rm au}$ and $n_{\rm pkts}=350$. The splitting required for Kubo regularisation was $N_{\rm kubo}=5$.